\documentclass[preprintnumbers,groupedaddress,nofootinbib,aps]{revtex4}

\usepackage{multirow}
\usepackage{amsmath,amssymb}
\usepackage{graphicx}
\usepackage{color}
\usepackage{hyperref}
\usepackage{url}

\begin{document}
\def\contentsname{{\normalsize Content}}
\def\tablename{Table}
\def\figurename{Figure}

\def\pveto{P_\text{veto}}
\def\nj{n_\text{jets}}
\def\meff{m_\text{eff}}
\def\ptmin{p_T^\text{min}}
\def\gtot{\Gamma_\text{tot}}
\def\as{\alpha_s}
\def\az{\alpha_0}
\def\gz{g_0}
\def\w{\vec{w}}
\def\sdag{\Sigma^{\dag}}
\def\s{\Sigma}
\newcommand{\psib}{\overline{\psi}}
\newcommand{\Psib}{\overline{\Psi}}
\newcommand\one{\leavevmode\hbox{\small1\normalsize\kern-.33em1}}
\newcommand{\Mpl}{M_\mathrm{Pl}}
\newcommand{\p}{\partial}
\newcommand{\mat}{\mathcal{M}}
\newcommand{\lag}{\mathcal{L}}
\newcommand{\ord}{\mathcal{O}}
\newcommand{\ope}{\mathcal{O}}
\newcommand{\qqquad}{\qquad \qquad}
\newcommand{\qqqquad}{\qquad \qquad \qquad}

\newcommand{\qb}{\bar{q}}
\newcommand{\matx}{|\mathcal{M}|^2}
\newcommand{\really}{\stackrel{!}{=}}
\newcommand{\msbar}{\overline{\text{MS}}}
\newcommand{\qns}{f_q^\text{NS}}
\newcommand{\lqcd}{\Lambda_\text{QCD}}
\newcommand{\met}{\slashchar{p}_T}
\newcommand{\pmiss}{\slashchar{\vec{p}}_T}

\newcommand{\sq}{\tilde{q}}
\newcommand{\go}{\tilde{g}}
\newcommand{\st}[1]{\tilde{t}_{#1}}
\newcommand{\stb}[1]{\tilde{t}_{#1}^*}
\newcommand{\nz}[1]{\tilde{\chi}_{#1}^0}
\newcommand{\cp}[1]{\tilde{\chi}_{#1}^+}
\newcommand{\cm}[1]{\tilde{\chi}_{#1}^-}
\newcommand{\CP}{CP}

\providecommand{\mg}{m_{\tilde{g}}}
\providecommand{\mst}{m_{\tilde{t}}}
\newcommand{\msn}[1]{m_{\tilde{\nu}_{#1}}}
\newcommand{\mch}[1]{m_{\tilde{\chi}^+_{#1}}}
\newcommand{\mne}[1]{m_{\tilde{\chi}^0_{#1}}}
\newcommand{\msb}[1]{m_{\tilde{b}_{#1}}}
\newcommand{\vsm}{\ensuremath{v_{\rm SM}}}

\newcommand{\mev}{{\ensuremath\rm MeV}}
\newcommand{\gev}{{\ensuremath\rm GeV}}
\newcommand{\tev}{{\ensuremath\rm TeV}}
\newcommand{\fb}{{\ensuremath\rm fb}}
\newcommand{\ab}{{\ensuremath\rm ab}}
\newcommand{\pb}{{\ensuremath\rm pb}}
\newcommand{\sign}{{\ensuremath\rm sign}}
\newcommand{\ifb}{{\ensuremath\rm fb^{-1}}}
\newcommand{\ipb}{{\ensuremath\rm pb^{-1}}}

\def\slashchar#1{\setbox0=\hbox{$#1$}           
   \dimen0=\wd0                                 
   \setbox1=\hbox{/} \dimen1=\wd1               
   \ifdim\dimen0>\dimen1                        
      \rlap{\hbox to \dimen0{\hfil/\hfil}}      
      #1                                        
   \else                                        
      \rlap{\hbox to \dimen1{\hfil$#1$\hfil}}   
      /                                         
   \fi}
\newcommand{\dslash}{\slashchar{\partial}}
\newcommand{\Dslash}{\slashchar{D}}

\newcommand{\eg}{\textsl{e.g.}\;}
\newcommand{\ie}{\textsl{i.e.}\;}
\newcommand{\etal}{\textsl{et al}\;}

\setlength{\floatsep}{0pt}
\setcounter{topnumber}{1}
\setcounter{bottomnumber}{1}
\setcounter{totalnumber}{1}
\renewcommand{\topfraction}{1.0}
\renewcommand{\bottomfraction}{1.0}
\renewcommand{\textfraction}{0.0}
\renewcommand{\thefootnote}{\fnsymbol{footnote}}

\newcommand{\rig}{\rightarrow}
\newcommand{\lrig}{\longrightarrow}
\renewcommand{\d}{{\mathrm{d}}}
\newcommand{\be}{\begin{eqnarray*}}
\newcommand{\ee}{\end{eqnarray*}}
\newcommand{\gl}[1]{(\ref{#1})}
\newcommand{\ta}[2]{ \frac{ {\mathrm{d}} #1 } {{\mathrm{d}} #2}}
\newcommand{\bee}{\begin{eqnarray}}
\newcommand{\eee}{\end{eqnarray}}
\newcommand{\beeq}{\begin{equation}}
\newcommand{\eeeq}{\end{equation}}
\newcommand{\mc}{\mathcal}
\newcommand{\mr}{\mathrm}
\newcommand{\ep}{\varepsilon}
\newcommand{\emt}{$\times 10^{-3}$}
\newcommand{\emfo}{$\times 10^{-4}$}
\newcommand{\emfi}{$\times 10^{-5}$}

\newcommand{\revision}[1]{{\bf{}#1}}

\newcommand{\hzero}{h^0}
\newcommand{\Hzero}{H^0}
\newcommand{\Azero}{A^0}
\newcommand{\PHiggs}{H}
\newcommand{\PW}{W}
\newcommand{\PZ}{Z}

\newcommand{\sw}{\ensuremath{s_w}}
\newcommand{\cw}{\ensuremath{c_w}}
\newcommand{\swd}{\ensuremath{s^2_w}}
\newcommand{\cwd}{\ensuremath{c^2_w}}

\newcommand{\mhhd}{\ensuremath{m^2_{\Hzero}}}
\newcommand{\mhh}{\ensuremath{m_{\Hzero}}}
\newcommand{\mlhd}{\ensuremath{m^2_{\hzero}}}
\newcommand{\Mlh}{\ensuremath{m_{\hzero}}}
\newcommand{\mad}{\ensuremath{m^2_{\Azero}}}
\newcommand{\mhpd}{\ensuremath{m^2_{\PHiggs^{\pm}}}}
\newcommand{\mhp}{\ensuremath{m_{\PHiggs^{\pm}}}}

 \newcommand{\sa}{\ensuremath{\sin\alpha}}
 \newcommand{\ca}{\ensuremath{\cos\alpha}}
 \newcommand{\cad}{\ensuremath{\cos^2\alpha}}
 \newcommand{\sad}{\ensuremath{\sin^2\alpha}}
 \newcommand{\sbd}{\ensuremath{\sin^2\beta}}
 \newcommand{\cbd}{\ensuremath{\cos^2\beta}}
 \newcommand{\cb}{\ensuremath{\cos\beta}}
 \renewcommand{\sb}{\ensuremath{\sin\beta}}
 \newcommand{\tanbd}{\ensuremath{\tan^2\beta}}
 \newcommand{\cotbd}{\ensuremath{\cot^2\beta}}
 \newcommand{\tanb}{\ensuremath{\tan\beta}}
 \newcommand{\tb}{\ensuremath{\tan\beta}}
 \newcommand{\cotb}{\ensuremath{\cot\beta}}

\newcommand{\GeV}{\ensuremath{\rm GeV}}
\newcommand{\MeV}{\ensuremath{\rm MeV}}
\newcommand{\TeV}{\ensuremath{\rm TeV}}

\title{Resolving the Higgs--Gluon Coupling with Jets}

\preprint{IPPP/14/50}
\preprint{DCPT/14/100}

\author{Malte Buschmann$^{1,2}$, 
        Christoph Englert$^3$, 
        Dorival Gon\c{c}alves$^2$, 
        Tilman Plehn$^1$, 
        Michael Spannowsky$^2$}

\affiliation{$^1$ Institut f\"ur Theoretische Physik, Universit\"at Heidelberg, Germany}
\affiliation{$^2$ Institute for Particle Physics Phenomenology, Department
  of Physics, Durham University, United Kingdom}
\affiliation{$^3$ SUPA, School of Physics and Astronomy, University of
  Glasgow, United Kingdom}

\begin{abstract}
  In the Standard Model the Higgs coupling to gluons is almost
  entirely induced by top quark loops. We derive the logarithmic
  structure of Higgs production in association with two jets. Just
  like in the one-jet case the transverse momentum distributions
  exhibit logarithms of the top quark mass and can be used to test the
  nature of the loop--induced Higgs coupling to gluons. Using Higgs
  decays to W bosons and to tau leptons we show how the corresponding
  analyses hugely benefit from the second jet in the relevant signal
  rate as well as in the background rejection.
\end{abstract}

\maketitle

\bigskip \bigskip \bigskip
\tableofcontents

\newpage

\section{Introduction}
\label{sec:intro}

After the recent discovery of a light, narrow, and likely fundamental
Higgs boson~\cite{higgs,discovery}, one of the main tasks of the
upcoming LHC runs will be to study the properties of this new particle. An
interesting aspect of the Higgs discovery is that it largely relies on
higher dimensional Higgs interactions which in the Standard Model are
induced by loops of heavy quarks and gauge bosons. While this indirect
information on Higgs coupling structures is complemented by precise
tree--level information in the Higgs--gauge sector, our
understanding of Higgs couplings to fermions largely relies on these
loop effects.

This shortcoming is most obvious in our currently very limited and
model--dependent understanding of the top Yukawa
coupling~\cite{sfitter,couplings_ex,couplings_th,higgs_review}. A measurement of the top Yukawa
coupling from associated Higgs and top production with a proper
reconstruction of the heavy states will be challenging even in the
upcoming LHC run~\cite{tth,thj,th}. This limitation is in stark
contrast with our theoretical interest, where a measurement of the
large top Yukawa coupling is crucial to extrapolate our understanding
of the Higgs mechanism from LHC energy scales to more fundamental,
high energies~\cite{sm_only}. Beyond the Standard Model this large size of
the top Yukawa suggests that any new physics stabilizing the scalar
Higgs mass should include a top partner, which in turn can contribute
to the loop--induced Higgs couplings to gluons and
photons~\cite{bsm_review}.\bigskip

To disentangle the Standard Model contribution for example to the
Higgs--gluon coupling from new physics effects we can use a particular
feature of the Standard Model loops: in the presence of a Yukawa
coupling the associated dimension-6 operators no longer
decouple. Instead, they induce a dimension-6 operator with a coupling
strengths which approaches a finite value in the limit of large top
masses. In this low energy limit the interactions between any number
of gluons and any number of Higgs bosons is given by a simple
effective Lagrangian~\cite{low_energy,lecture}. While this
approximation provides a very good prediction of the inclusive Higgs
production rate it leads to $\ord(10\%)$ deviations in most
distributions for the $gg \to H$ production
process~\cite{spirix_nlo,robert,spirix_review} and fails quite
spectacularly for Higgs pair production~\cite{higgs_pair}. Turning
this argument around, we can use kinematic distributions in Higgs
production processes to test our assumption that the Higgs--gluon
interactions are induced by heavy quarks.

Physics beyond the Standard Model might also exhibit non--decoupling
effects in the effective Higgs couplings. One such example is a fourth
generation of chiral fermions, where the effects from new physics are
of the same size as the Standard Model prediction. Because they are
not described by a small parameter such scenarios are largely ruled
out altogether. In new physics extensions which do decouple, the
characteristic small parameter is typically the ratio of the
electroweak scale to the new physics mass scale. This mass ratio is
constrained to be below $\ope(1/10)$, with a possible exception of
supersymmetric top partners which are experimentally still allowed to
reside around the top mass scale~\cite{light_stop}. Under this
assumption of heavy new states the low energy approximation to the
Higgs--gluon couplings holds for the loop contributions from physics
beyond the Standard Model~\cite{keith,uli}. This makes it
straightforward to interpret deviations from kinematic features
predicted for the heavy quark loops in terms of new physics
scenarios~\cite{schlaffer_spannowsky}.\bigskip

The key question in the above reasoning is which kinematic features
are best suited to test the heavy quark origin of the Higgs--gluon
couplings.  It has been known for a long time that the transverse
momentum distribution of Higgs production in association with a hard
jet exhibits a logarithmic dependence on the top
mass~\cite{keith,uli}. Recently, this effect has been proposed as a
handle to test the Standard Model assumption that the Higgs--gluon
coupling is exclusively due to heavy quark
loops~\cite{sanz,azatov,andi,englert_spannowsky}.

In this paper we for the first time go beyond Higgs production with a
single hard jet. Higgs production in gluon fusion associated with two
hard initial state radiation jets offers a much richer set of
kinematic distributions\footnote{We will refer to this process as
  vector boson fusion (VBF) and neglect the numerically small weak
  boson fusion contributions. Moreover, we do not require the usual
  forward tagging jets, but two hard jets defining the hard process
  together with the Higgs.}. It is well known that the correlations of
the two initial state radiation jets reflect the higher dimensional
structure of the Higgs coupling to gluons or any other hard
process~\cite{phi_jj}. In this study we will use the two hard jets to
extract the top mass dependence of the Higgs--gluon coupling.

First, we will show that the logarithmic top mass dependence in the
VBF topology is the same as for Higgs production with a single
jet. Adding a second hard jet to the hard process \cite{dieter} shifts
a sizeable number of Higgs events from phase space regions which are
not sensitive to top mass effects to regions which are sensitive. We
will find that the sensitivity of the VBF topology to top mass effects
should exceed the sensitivity of the Higgs--plus--one--jet
channel. Moreover, the VBF topology allows for a much improved
background suppression in the $H \to \tau \tau$ and $H \to WW$
channels. This way, a second hard jet is not just an improvement of a
dominant 1-jet analysis; the 2-jet hard process is more sensitive
to top mass effects, the correlations of the second hard jet and the
logarithmic top mass dependence are not covered by a parton shower
description, and the second hard jet makes a big difference in the
background rejection.

\section{Top mass effects}
\label{sec:logs}

The main production process responsible for the Higgs discovery is
gluon fusion, mediated by the Higgs coupling to a pair of gluons. This
interaction does not exist at tree level, {\sl i.e.}, as part of the
renormalizable dimension-4 Lagrangian. It is induced by heavy quarks,
in the Standard Model dominantly via top quark
loop~\cite{low_energy,lecture},
\begin{alignat}{5}
\lag_{ggH} &\supset \;
g_{ggH} \; \frac{H}{v} \;  \, G^{\mu\nu}G_{\mu\nu} \notag \\
\frac{g_{ggH}}{v} 
 &=
-i \,\frac{\alpha_s}{8 \pi} \; 
   \frac{1}{v} \; \tau \left[1+(1-\tau)f(\tau)\right] 
\qqqquad 
f(\tau) 
\stackrel{\text{on-shell}}{=} 
\left( \arcsin \sqrt{\dfrac{1}{\tau}}\right)^2 
\stackrel{\tau \to \infty}{=} 
\frac{1}{\tau} + \frac{1}{3\tau^2} + \ope \left( \frac{1}{\tau^3} \right) 
\; ,
\label{eq:higgs_eff1}
\end{alignat}
all in terms of $\tau = 4m_t^2/m_H^2 > 1$. Barring prefactors the
function $f$ corresponds to the scalar three--point function for a
closed top loop. In the usual kinematic configuration for single Higgs
production the coupling $g_{ggH}$ depends only on the top and Higgs
masses, as indicated above. Once it appears as part of a more complex
Feynman diagram the coupling $g_{ggH}$ will depend on the momenta of
all three external states as well as on the top mass. This will become
our main reason to define the hard process including two hard jets
rather than one jet plus a parton shower.

\begin{figure}[!b]
 \includegraphics[width=0.6\textwidth]{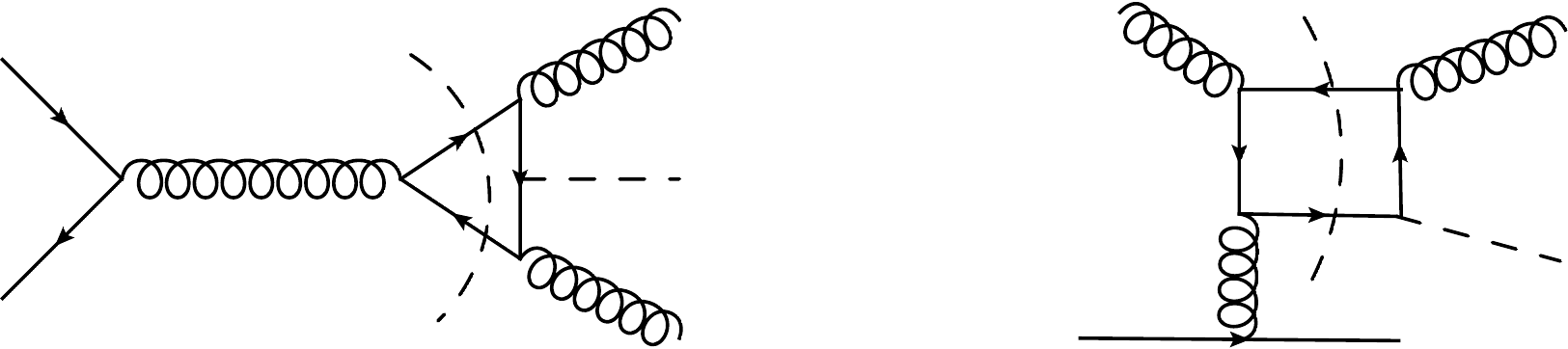}
\caption{Sample Feynman diagrams for the processes $qq\rightarrow Hgg$
  and $gq\rightarrow Hgq$, indicating the cuts which contribute to
  absorptive parts.}
\label{fig:feyn_absorptive} 
\end{figure}

In the simple low energy limit the interaction vertices between any
number of gluons and any number of Higgs bosons can be described by
the Lagrangian
\begin{alignat}{5}
\lag_{ggH} 
=\frac{\alpha_s}{12\pi} \;  
 \log \left(1+\frac{H}{v} \right) \;  
 G^{\mu\nu}G_{\mu\nu} \;
\supset \frac{\alpha_s}{12 \pi} \; 
 \frac{H}{v} \; 
 G^{\mu\nu}G_{\mu\nu} \; .
\label{eq:higgs_eff2}
\end{alignat}
The top Yukawa coupling in the top loop violates the
decoupling theorem, so the interaction approaches a finite limit
\cite{low_energy}. This non--decoupling property in combination with
the absence of a dimension-4 Higgs coupling to gluons is unique to the
dimension-6 operators mediating the Higgs couplings to gluons and
photons, which are to a large degree responsible for the Higgs
discovery~\cite{discovery}.\bigskip

One question which we have to answer based on LHC measurements is if
the top Yukawa coupling is indeed responsible for the observed
Higgs--gluon coupling, or if other top partners contribute to the
corresponding dimension-6 operator. In two different conventions the
relevant part of the Higgs interaction Lagrangian including a finite
top mass and free couplings reads
\begin{alignat}{5}
\lag_\text{int} &\supset
\left[  \kappa_t \; g_{ggH} + \kappa_g\frac{\alpha_s}{12 \pi}  \right] \; 
\frac{H}{v} \; 
G_{\mu\nu}G^{\mu\nu}
- \kappa_t \; 
\frac{m_t}{v} H \left( \bar{t}_R t_L + \text{h.c.} \right)
&&\text{Refs.~\cite{andi}}  \notag \\
&=
\left( 1 + \Delta_t + \Delta_g \right) \; g_{ggH} \; 
\frac{H}{v} \; 
G_{\mu\nu}G^{\mu\nu}
- (1 + \Delta_t ) \; 
\frac{m_t}{v} H \left( \bar{t}_R t_L + \text{h.c.} \right) 
\qqquad &&\text{\textsc{SFitter}~\cite{sfitter}} \; .
\label{eq:lagrangian}
\end{alignat}
We show the \textsc{SFitter} conventions to indicate that the parameters
$\kappa_t$ and $\kappa_g$ are indeed part of the usual LHC coupling analyses.
The new aspect is to extract them from distributions rather than rates.
As alluded to above, the dimension-6 operator is defined not only
without any reference to the top mass, but also with the entire
momentum dependence arising from the gluon field strengths.  One
physics scenario which could serve as an ultraviolet extension of
Eq.\eqref{eq:lagrangian} would be the Standard Model with an extended
Higgs sector and an unobserved top
partner~\cite{sfitter,schlaffer_spannowsky}. Throughout this paper we will rely on two
reference points unless otherwise mentioned,
\begin{alignat}{5}
(\kappa_t,\kappa_g)_\text{SM}=(1,0) 
\qqquad \text{and} \qqquad 
(\kappa_t,\kappa_g)_\text{BSM}=(0.8,0.2) \; .
\label{eq:points}
\end{alignat}
In the second point the contributions from a top partner to a good
approximation compensate for the reduced top Yukawa in the
Higgs--gluon coupling, leaving the observed Higgs cross section at the
LHC unchanged.

\subsection*{Absorptive terms}

\begin{figure}[!t]
 \includegraphics[width=0.4\textwidth]{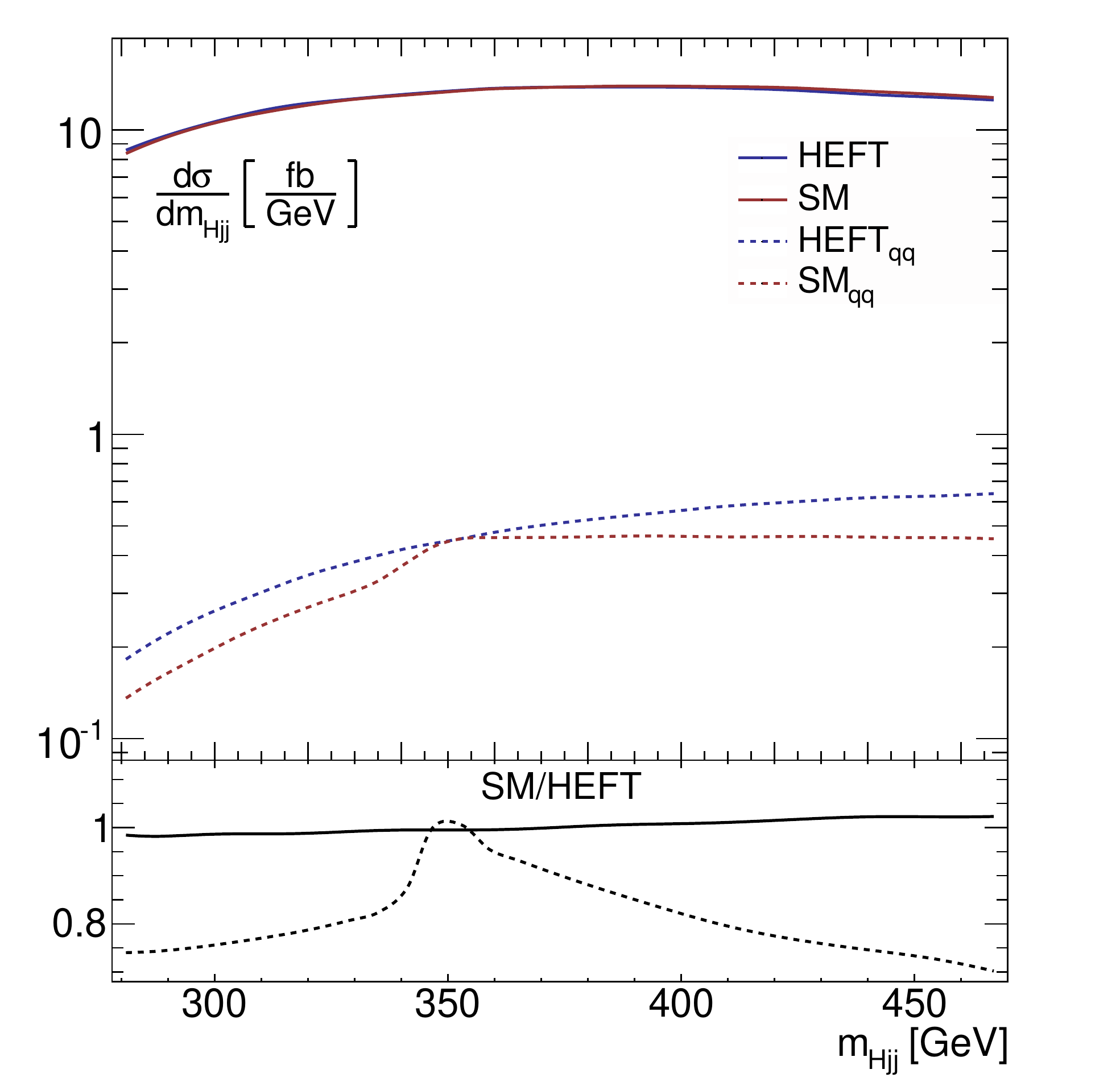}
 \hspace*{0.1\textwidth}
 \includegraphics[width=0.4\textwidth]{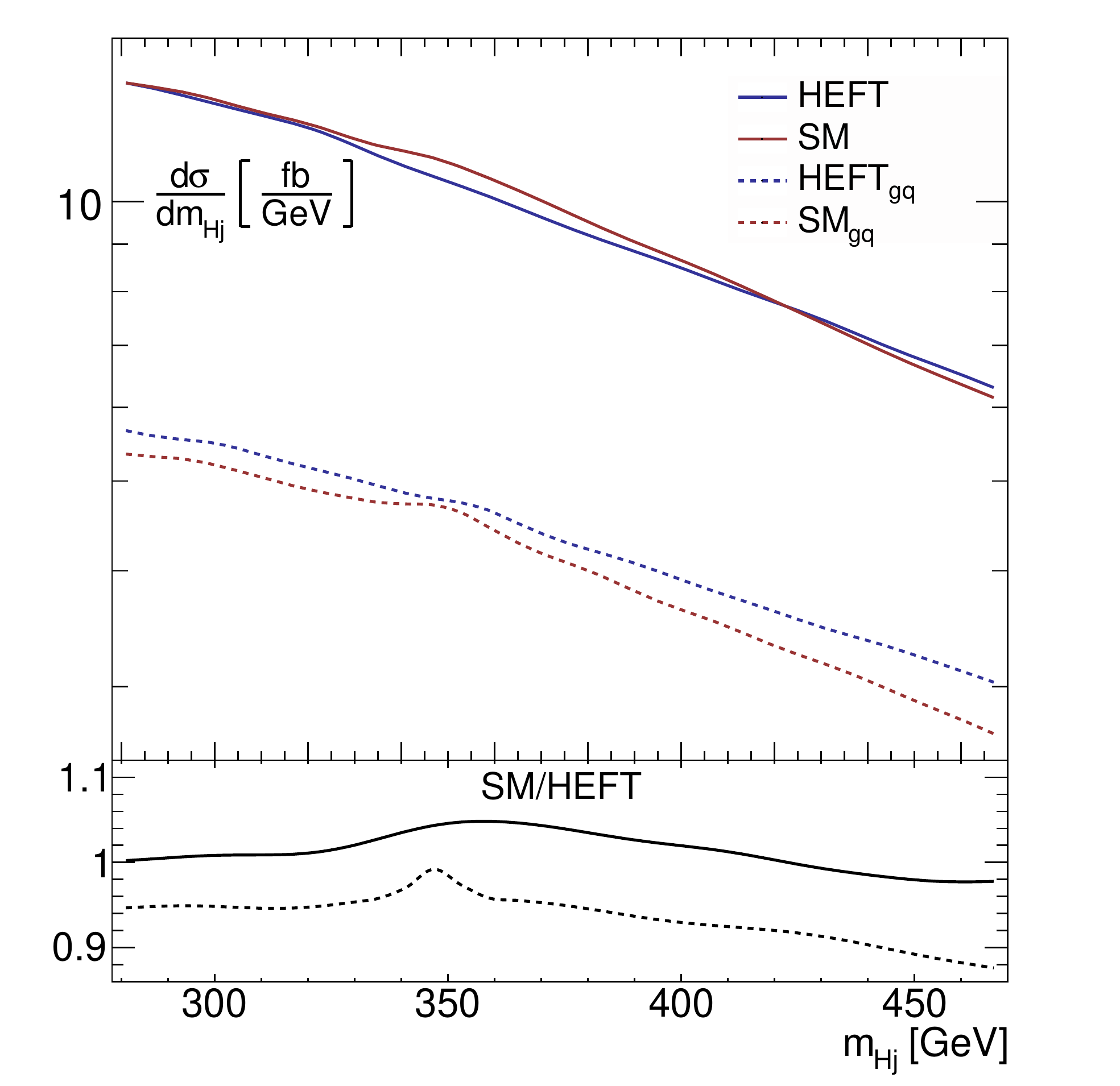}
\caption{Differential distributions for $m_{Hjj}$ (left) and for
  $m_{Hj}$ (right) for the $Hjj$ process. The Standard Model
  curves (SM) include the full top mass dependence while the low
  energy effective field theory approximation (HEFT) relies on the
  approximation in Eq.\eqref{eq:higgs_eff2}. The index `qq' (`gq') indicates
  Feynman diagrams with an incoming quark pair (gluon-quark). We assume $\sqrt{S} =
  13$~TeV.}
\label{fig:absorptive} 
\end{figure}

Absorptive terms in the top loop inducing the effective Higgs--gluon
coupling are well known from the behavior of the cross section as a
function of the (formerly unknown) Higgs
mass~\cite{lecture,spirix_review}. At $m_H = 2 m_t$ the formula for the
scalar integral given in Eq.\eqref{eq:higgs_eff1} develops an
imaginary absorptive part, leading to a kink in the LHC cross
section. Given the now fixed Higgs mass of 126~GeV the question is how
we can search for such effects at the LHC. For example, in Higgs
production in association with two jets the same absorptive parts
should appear in the loop integrals shown in
Fig.~\ref{fig:feyn_absorptive},
\begin{alignat}{5}
m_{Hg} = 2 m_t 
\qqquad \text{and} \qquad 
m_{Hgg} = 2 m_t  \; .
\label{eq:kinks}
\end{alignat}
To illustrate the size of such absorptive effects we study the process
$pp \to Hjj$ at parton level in Fig.~\ref{fig:absorptive}. It includes
the loop--induced $gggH$ interaction which indeed shows an absorptive
part around $m_{Hj} \sim 350$~GeV, as indicated in
Eq.\eqref{eq:kinks}.  We see that these absorptive parts are very
small for both distributions and will hardly allow us to make a
qualitative statement about the origin of the effective Higgs--gluon
coupling, not even talking about a measurement of $\kappa_t$ and
$\kappa_g$.

\subsection*{Top--induced logarithms}

Higgs production in association with a hard jet probes a logarithmic
top mass dependence of the loop--induced
coupling~\cite{keith,uli}. This effect has recently been transformed
into a proposed experimental separation of the coupling modifications
$\kappa_t$ and $\kappa_g$ in this production
channel~\cite{sanz,azatov,andi,schlaffer_spannowsky}. In the high
energy limit, or for small top and Higgs masses, the leading term of
the matrix element for the partonic subprocess $gg \to Hg$ scales like
\begin{alignat}{5}
|\mat_{Hj}|^2 \propto m_t^4 \; \log^4 \frac{p_T^2}{m_t^2} \; .
\label{eq:uli_log}
\end{alignat}
The transverse momentum constitutes the external energy scale in the
limit of $p_T \gg m_H, m_t$. If the effective Higgs--gluon coupling is
not induced by the top quark this logarithm does not occur.\bigskip

Next, we look at the logarithmic structure for the more complex final
state of Higgs production in association with two jets.  In the
presence of several external mass scales it is not clear which
final--state invariant drives the logarithmic top mass dependence. The
simplest subprocess $q \bar{q} \to q \bar{q} H$ only probes the
effective $ggH$ coupling, but with two off-shell gluons at sizeable
virtualities. In terms of the virtualities of $Q_{1,2} > 0$ of the
space--like or $t$-channel gluons the corresponding scalar three point
function scales like
\begin{alignat}{5}
|\mat_{Hjj}|^2 \propto 
\frac{m_t^4}{( Q_1^2 - Q_2^2 )^2} 
\left( \log^2 \frac{Q_1^2}{m_t^2}
     - \log^2 \frac{Q_2^2}{m_t^2}
\right)^2
\stackrel{Q_1 \gg Q_2}{=} 
\frac{m_t^4}{Q_1^4} 
\; \log^4 \frac{Q_1^2}{m_t^2} \; .
\label{eq:our_log_1}
\end{alignat}
In the collinear limit the virtuality of the incoming parton splitting
is linked to the transverse momentum of the forward tagging jet
through a simple linear transformation. Logarithms in the virtuality
can be directly translated into logarithms of the transverse momentum,
independent if they are scaling logarithms which get absorbed into the
parton densities or if they affect the hard process~\cite{boos}.

\begin{figure}[t!]
  \includegraphics[width=0.24\textwidth]{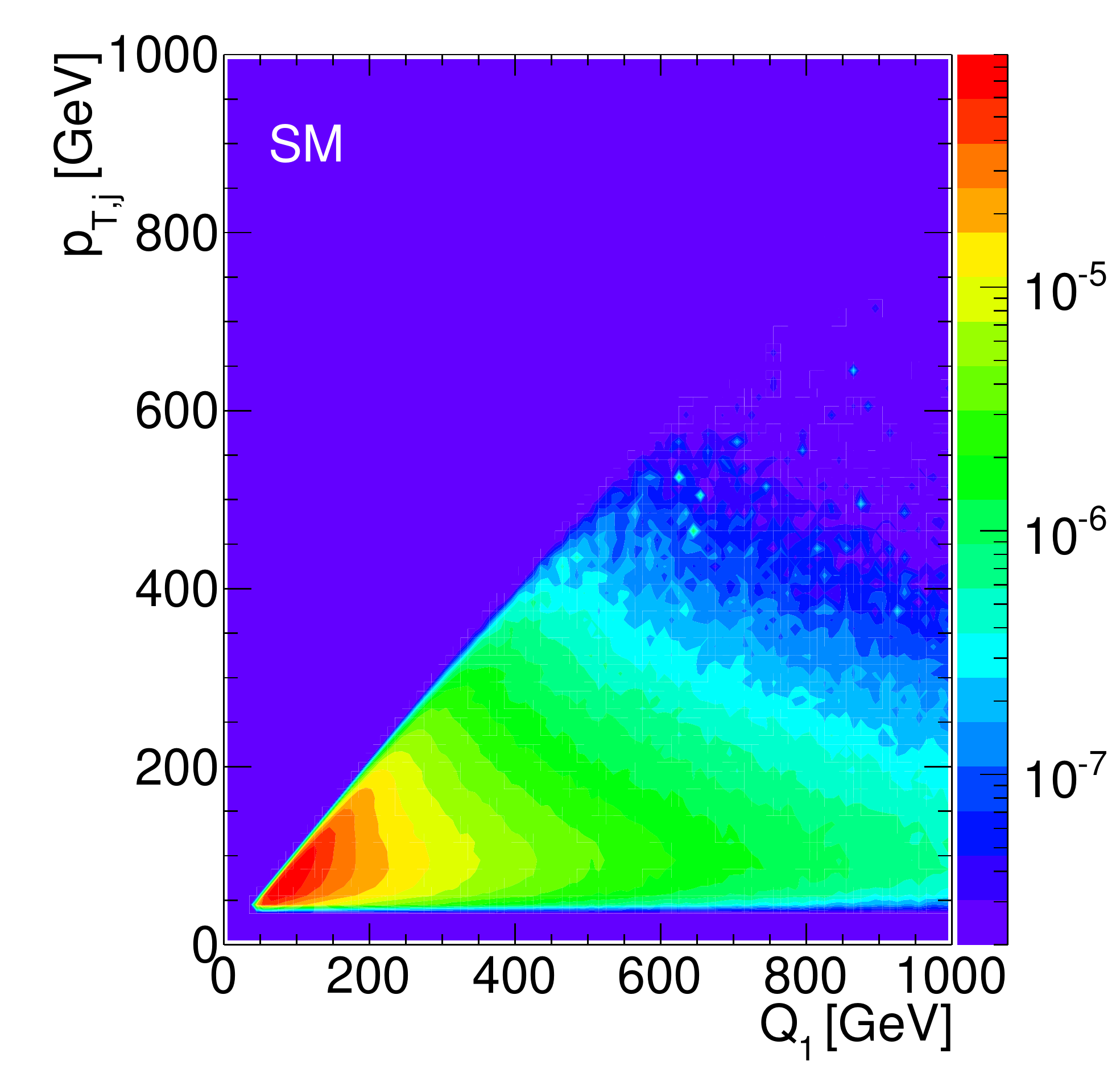}
   \hspace*{0.0\textwidth}
  \includegraphics[width=0.24\textwidth]{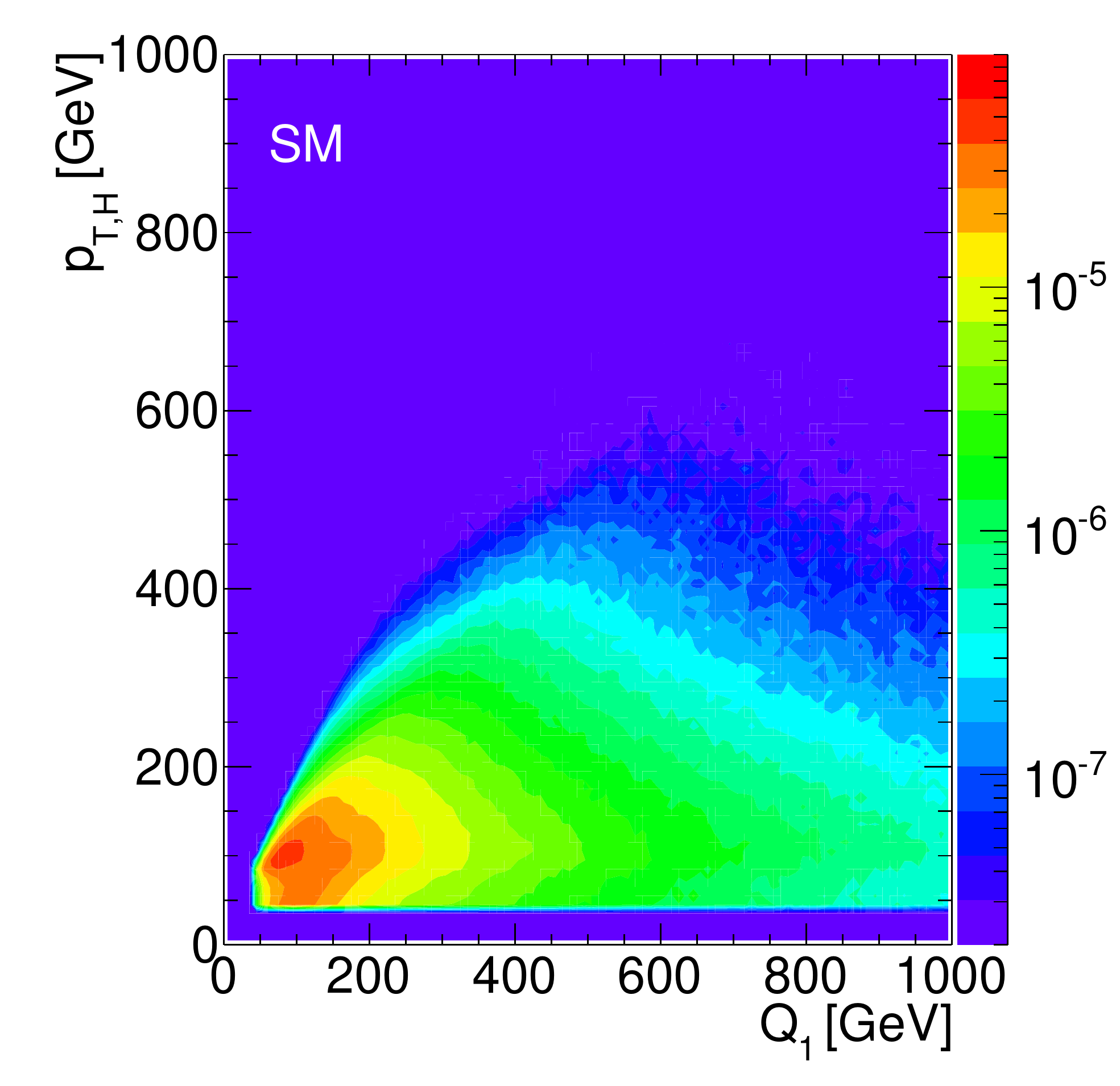}
   \hspace*{0.0\textwidth}
  \includegraphics[width=0.24\textwidth]{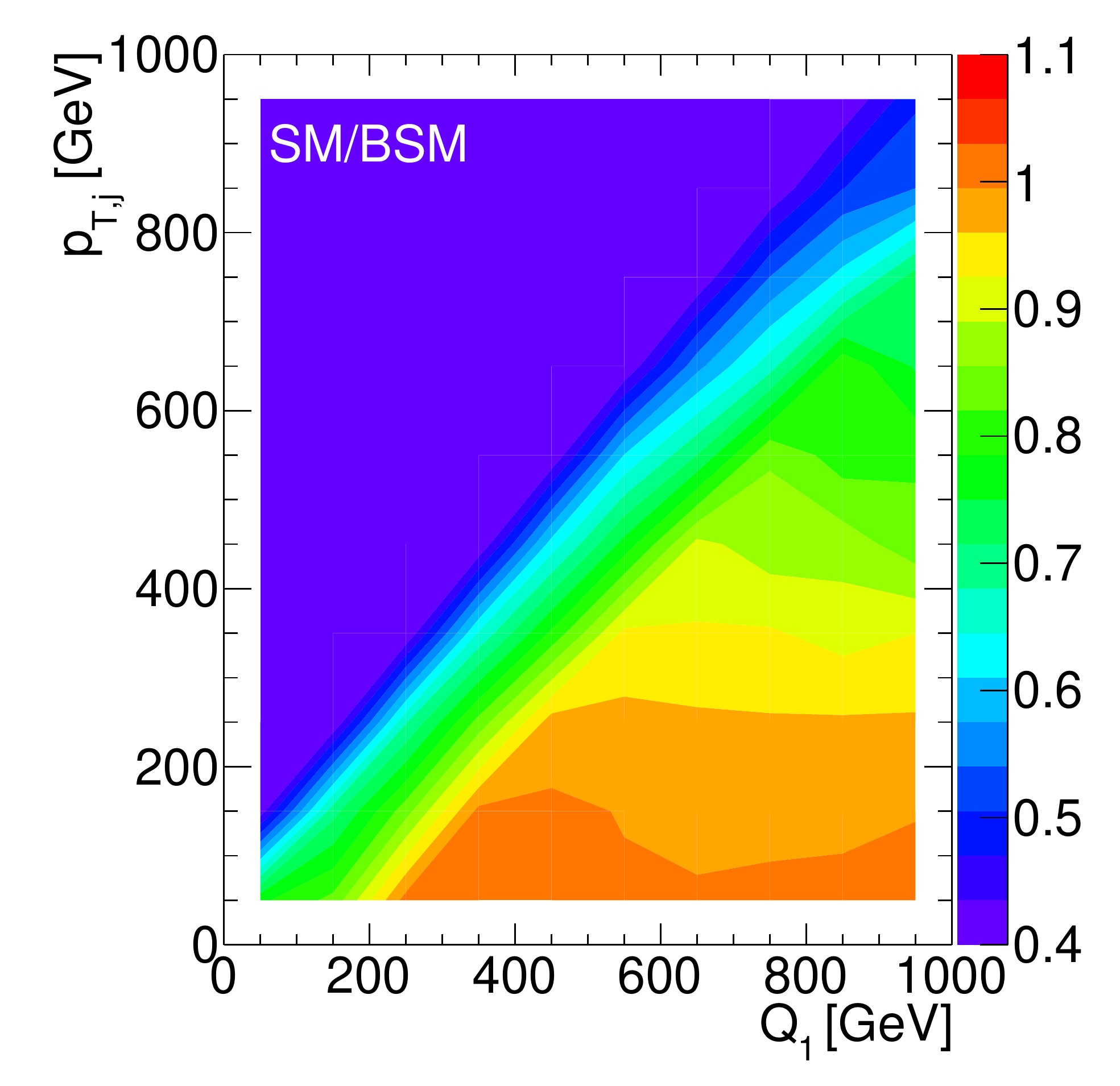}
   \hspace*{0.0\textwidth}
  \includegraphics[width=0.24\textwidth]{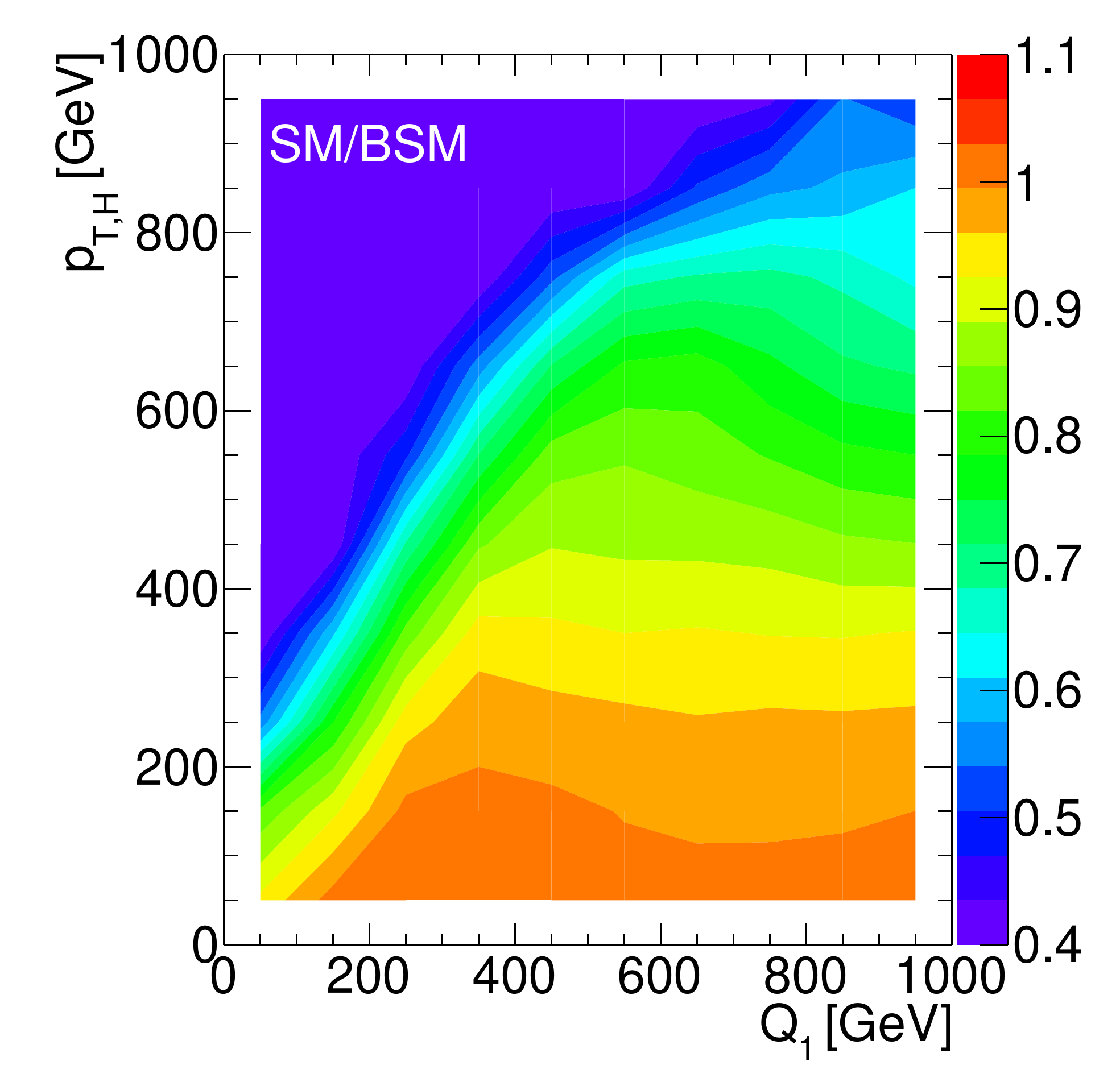}
\caption{Left to right: correlation plots for the leading $p_{T,j}$ vs $Q_1$
  and $p_{T,H}$ vs $Q_1$ for $H jj$ production in the Standard
  Model, $\kappa_{t,g}=(1,0)$. We also show the ratio SM/BSM, where
  BSM is defined as $\kappa_{t,g}=(0.8,0.2)$.}
\label{fig:ptj_q1} 
\end{figure}

In the limit of one significantly harder tagging jet $Q_1 \gg Q_2$
recoiling against the Higgs boson the diagrams in the vector boson
fusion topology scale like
\begin{alignat}{5}
|\mat_{Hjj}|^2 
\propto 
m_t^4 \; \log^4 \frac{p_{T,j}^2}{m_t^2} 
\sim
m_t^4 \; \log^4 \frac{p_{T,H}^2}{m_t^2} \; .
\label{eq:our_log_2}
\end{alignat}
In this step we assume a linear relation between the virtuality and
the transverse momentum of the additional jets~\cite{boos}.  In the
left panel of Fig.~\ref{fig:ptj_q1} we show the correlation between
the leading $p_{T,j}$ and the corresponding gluon virtuality for the
SM hypothesis and clearly see the expected correlation with $p_{T,j} >
Q$.  Away from the diagonal we only find events with $p_{T,j_1} <
Q_1$, in agreement with the kinematic considerations of
Ref.~\cite{boos}. This pattern gets transferred to the transverse
momentum of the recoiling Higgs. In the right two panels we show the
same kinematic correlation for the ratio SM/BSM.  We see the same
increase of the dimension-6 operators at larger transverse momenta as
in the $Hj$ channel~\cite{sanz,andi,azatov,schlaffer_spannowsky}. For
given $p_{T,j_1}$ values this ratio is independent of the virtuality.
This means that while the virtuality is fixed by the steep gluonic
parton densities the top mass logarithm feeds on the transverse
momentum and the jet momentum in the beam direction.\bigskip

\begin{figure}[!b]
 \includegraphics[width=0.4\textwidth]{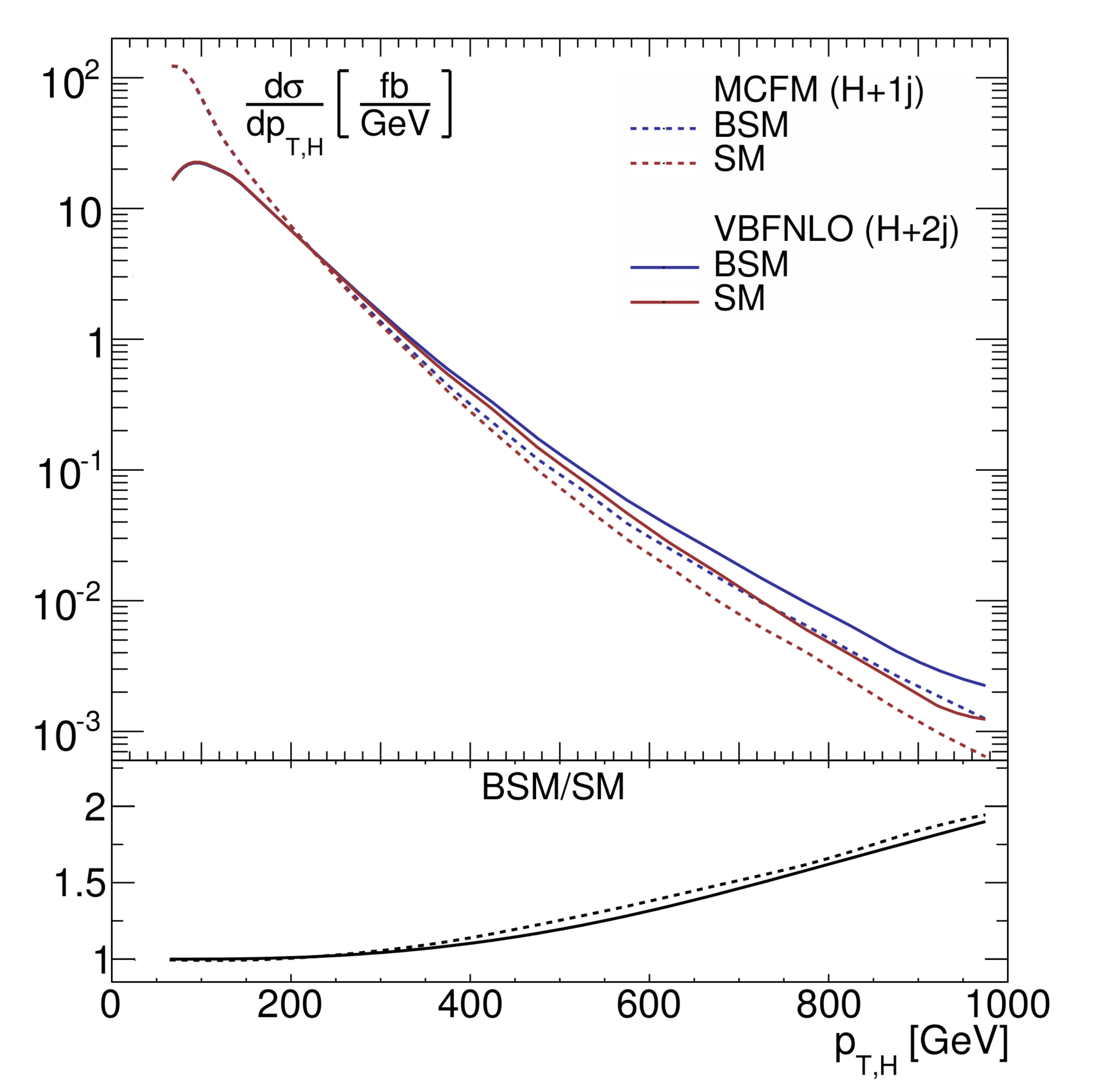}
  \hspace*{0.1\textwidth}
 \includegraphics[width=0.4\textwidth]{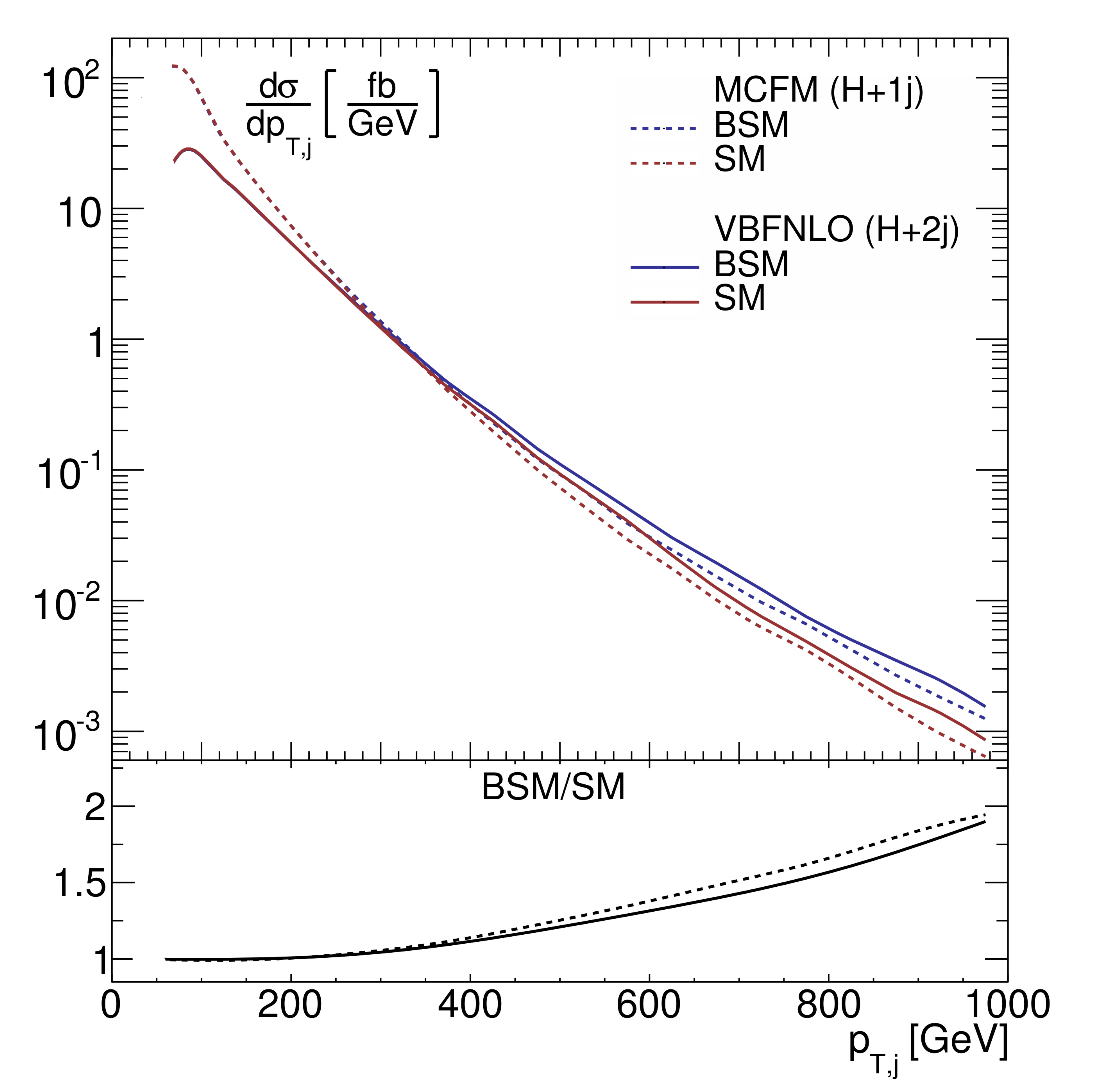}
\caption{Parton--level $p_{T,H}$ (left) and $p_{T,j_1}$ (right)
  distributions for $Hj$ and $Hjj$ production. The red curve
  corresponds to the Standard Model $\kappa_{t,g}=(1,0)$, while the
  blue curves follow from the BSM hypothesis
  $\kappa_{t,g}=(0.8,0.2)$. We assume $\sqrt{S}=13$~TeV.}
\label{fig:partonic_log} 
\end{figure}

After ensuring that the top mass logarithms in $Hj$ and $Hjj$ have the
same origin we can compare their numerical impact.  In
Fig.~\ref{fig:partonic_log} we show the dependence of the $Hj$ and
$Hjj$ production rates on the transverse momentum of the leading
tagging jet and the Higgs, based on the \textsc{Mcfm}~\cite{mcfm} and
\textsc{Vbfnlo}~\cite{vbfnlo} implementations. We have validated this
modified \textsc{Mcfm} dimension-6 setup against an independent
implementation based on \textsc{Vbfnlo}. We compare the prediction of
the Standard Model $\kappa_{t,g} = (1,0)$, to an additional BSM
contribution from the dimension-6 operator $\kappa_{t,g} =(0.8,0.2)$,
as defined in Eq.\eqref{eq:points}.  For both channels there appears a
logarithmic enhancement for transverse momenta larger than twice the
top mass.

The full $Hjj$ production process includes one--loop triangle, box and
pentagon contributions, which cannot be separated. However, the
different $qq$, $gq$ and $gg$ initial states offer a handle to
determine the size of their relative contributions.  For the $qq$ and $gq$
initial states we have triangle and box diagrams, and the $gg$ initial state
will include pentagons. For all initial states we find an enhanced 
dimension-6 BSM component at large Higgs and jet transverse momenta. 
The effect is strongest for incoming quarks and less pronounced for pure gluon
amplitudes. This confirms our original assumption that the top mass
logarithm arises from the VBF topology with an effective triangular
$ggH$ interaction for all initial states. This topology is
approximately added to the $Hj$ simulation once we include a parton
shower to simulate initial state radiation. However, if both 
jets are hard the VBF topology is correctly described by the
appropriate hard process, which includes the Higgs as well as two jets.\bigskip

The comparison of the $Hj$ and the $Hjj$ channels in
Fig.\ref{fig:partonic_log} also shows that for one recoiling jet most
of the cross section comes from phase space regions which do not
resolve the effective Higgs--gluon coupling. In comparison, for two
hard jets recoiling against the boosted Higgs the drop in the total
cross section appears exclusively in the insensitive regime, while
even in terms of absolute event numbers the sensitivity to the top
mass logarithm increases. If indeed the hard $Hjj$ process is
numerically more relevant in the high-$p_T$ regime than the hard $Hj$
process we need to worry about even more jets. We can only
speculate about this, but from the above observation that the top mass
logarithm arises from the VBF topology the third jet would be most
helpful if arising from a final state splitting. Such configurations
should be reasonably well described by the final state parton
shower.\bigskip

\begin{figure}[t!]
 \includegraphics[width=0.325\textwidth]{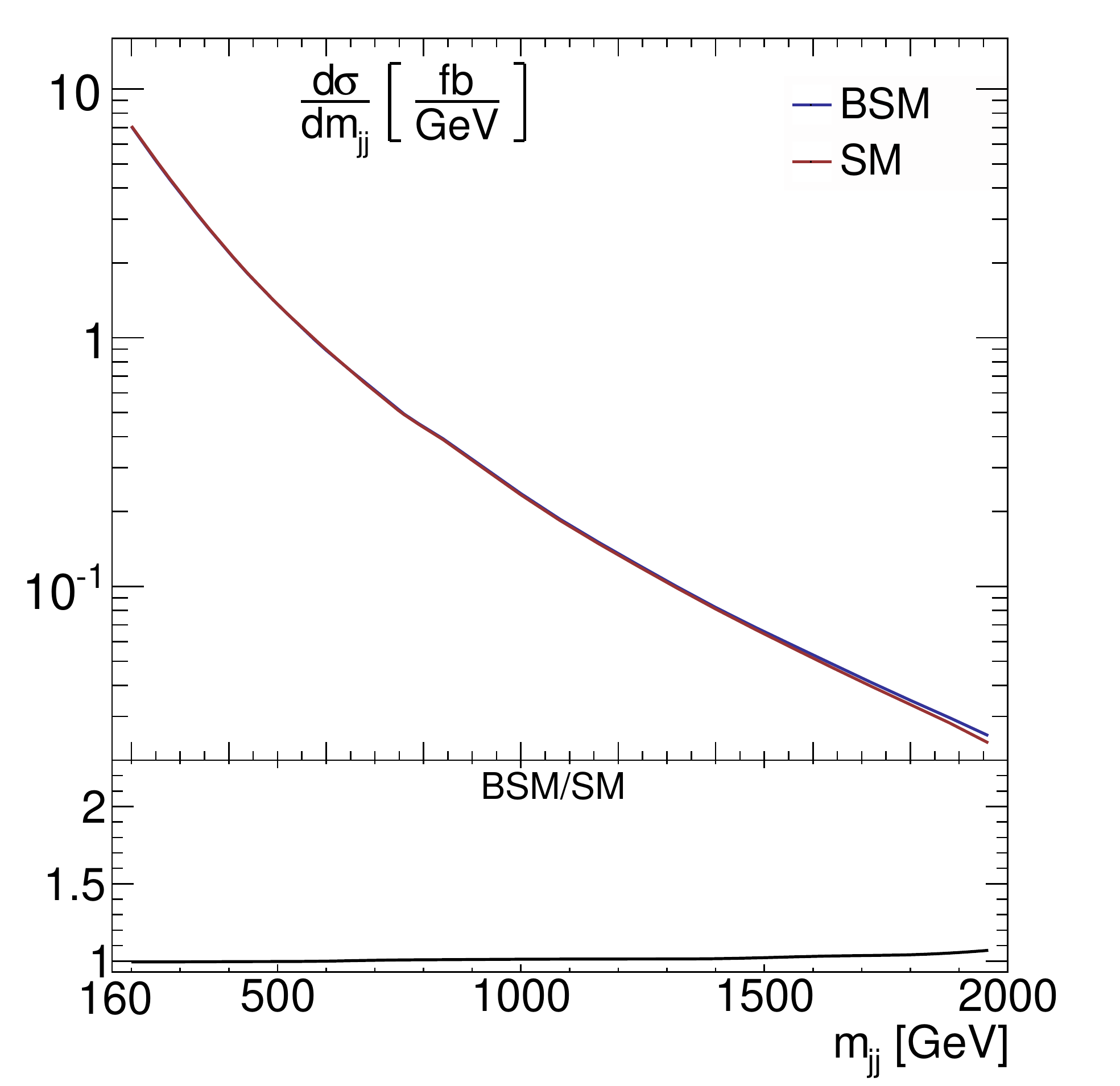}
 \includegraphics[width=0.325\textwidth]{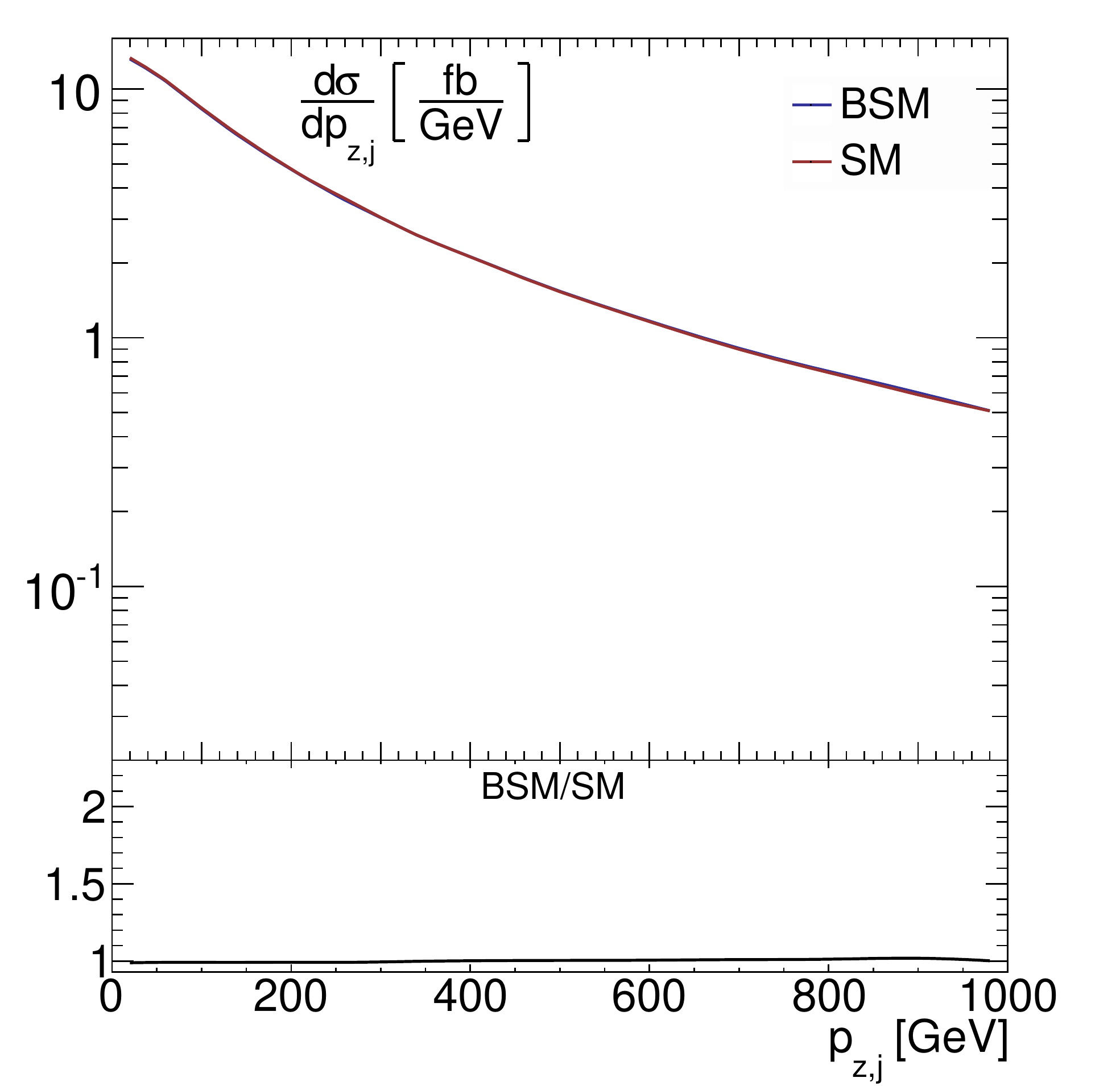}
 \includegraphics[width=0.325\textwidth]{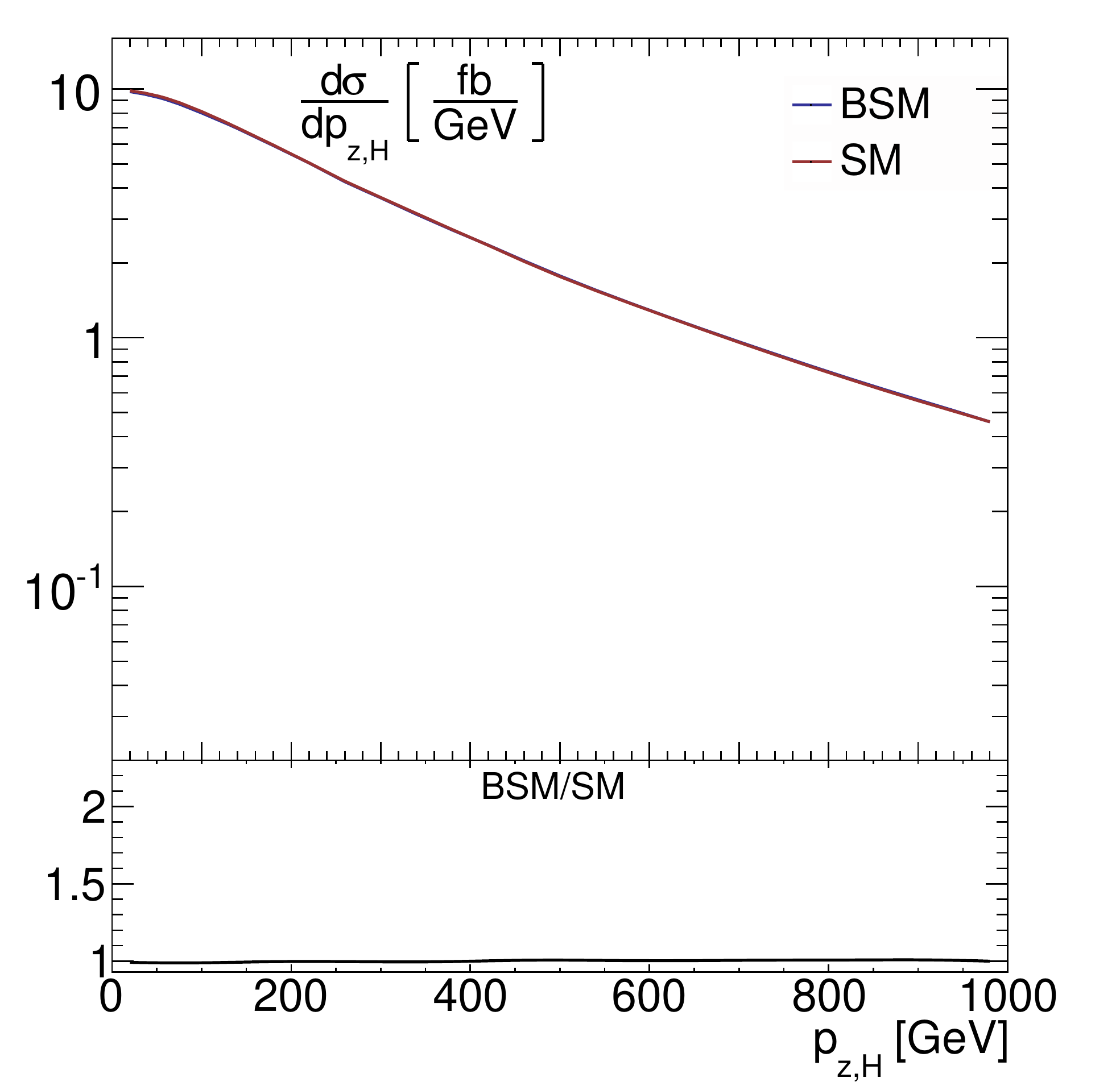}
\caption{Parton--level $m_{jj}$ (left), $p_{z,j_1}$ (center) and $p_{z,H}$
  (right) distributions for $Hjj$ production.}
\label{fig:partonic_nolog} 
\end{figure}

After isolating the top mass logarithm in the transverse momentum
spectra for $Hjj$ production given by Eq.\eqref{eq:our_log_2} we need
to sadly convince ourselves that there are no additional top
mass logarithms in this process. For example, there could be very
promising logarithms in the largest momentum scale, \ie $\log
m_{jj}/m_t$.  In Fig.~\ref{fig:partonic_nolog} we show the $m_{jj}$
distribution as well as the leading $p_{z,j_1}$ and $p_{z,H}$
distributions for $Hjj$ production. For the top--induced coupling and
the dimension-6 coupling they are perfectly aligned,
indicating that none of these observables are affected by top mass
logarithms.  The top mass dependence really only appears in the
transverse momentum spectra. In the following we will focus on the
transverse momentum of the Higgs, while eventually an experimental
analysis could include both, $p_{T,H}$ and the leading
$p_{T,j}$.

\subsection*{Including the interference}

\begin{figure}[!t]
 \includegraphics[width=0.4\textwidth]{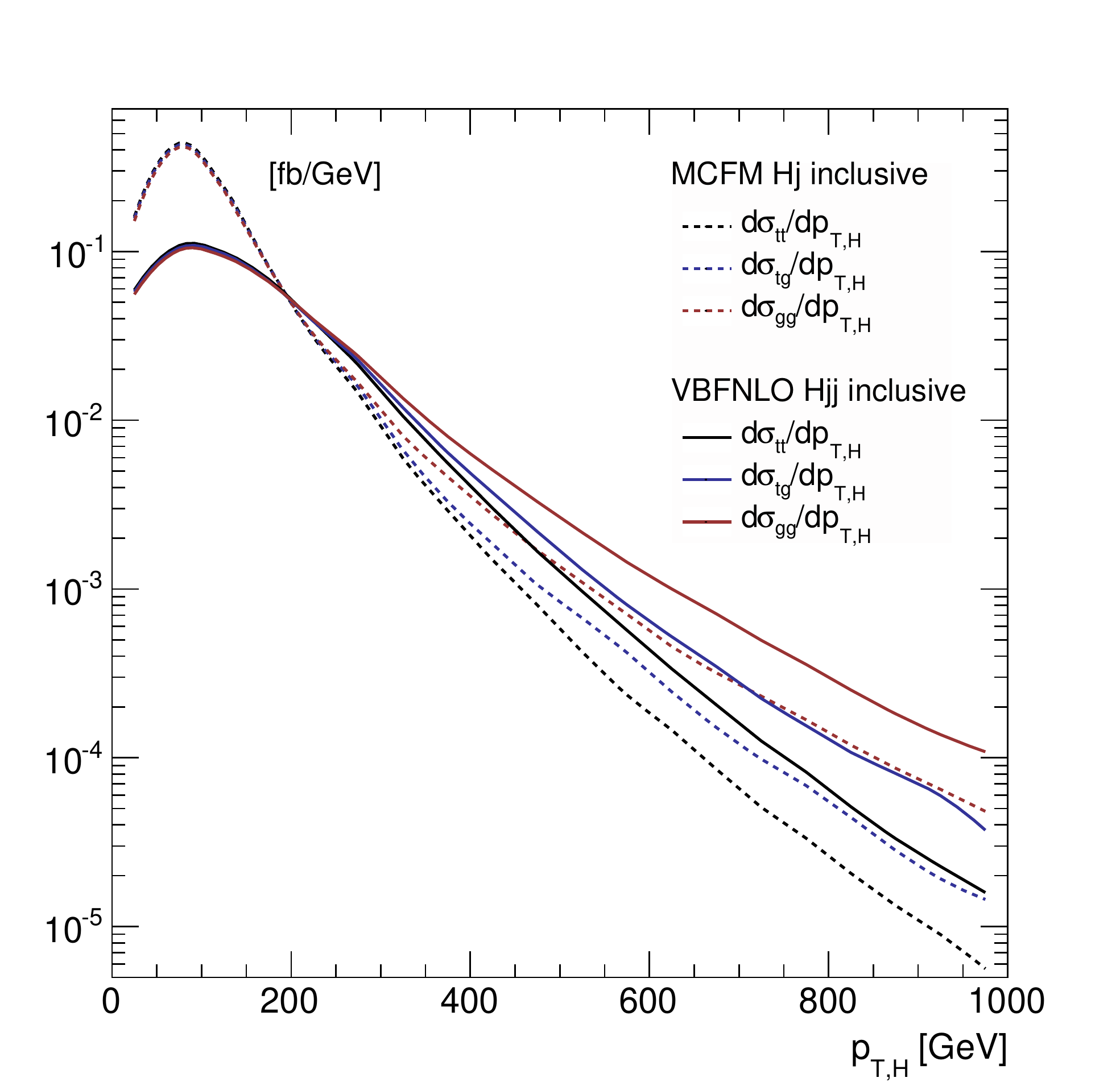}
\caption{Transverse momentum distribution for $Hj$ production (based
  on \textsc{Mcfm}) and $Hjj$ production (based on
  \textsc{Vbfnlo}). Both codes use \textsc{Pythia8} for the parton
  shower. The top--induced and dimension-6 contributions as well as
  their interference are defined in Eq.\eqref{eq:distribution}.  We
  assume $\sqrt{S}=13$~TeV and for technical reasons include a decay
  $H \to \tau \tau$ with minimal cuts.}
\label{fig:mother_distrib} 
\end{figure}

Based on the interaction Lagrangian in Eq.\eqref{eq:lagrangian} we can
easily translate the modified coupling strengths into differential or
total LHC cross sections. For simplicity, we keep all other
tree--level Higgs interactions unchanged, so the expected slight shift
in the photon--Higgs coupling will be of no phenomenological
relevance. The matrix element for Higgs production in gluon fusion is
based on the $H G_{\mu \nu} G^{\mu \nu}$ interaction and will consist
of two terms,
\begin{alignat}{5}
\mat = \kappa_t \mat_t + \kappa_g \mat_g \; ,
\label{eq:amplitude}
\end{alignat}
where the index $g$ indicates the dimension-6 operator
contribution and all prefactors except for the $\kappa_j$ are absorbed
in the definitions of $\mat_j$. For the matrix element squared and
any kinematic distributions this means
\begin{alignat}{5}
\frac{d\sigma}{d\ope} & 
= \kappa_t^2 \; \frac{d\sigma_{tt}}{d\ope}  
+ \kappa_t \kappa_g \; \frac{d\sigma_{tg}}{d\ope}  
+ \kappa_g^2 \; \frac{d\sigma_{gg}}{d\ope} \; ,
\label{eq:distribution}
\end{alignat}
where for small deviations from the Standard Model the last term will
be numerically irrelevant.  In Fig.~\ref{fig:mother_distrib} we
present the three transverse momentum distributions for the Higgs, on
which we will rely for the remaining analysis. To be consistent, we
use \textsc{Mcfm}+\textsc{Pythia8}~\cite{mcfm,pythia} for the hard
$Hj$ production process with the scale choice $\mu_F^2=\mu_R^2=m_H^2+
p_{T,j}^2$ and \textsc{Vbfnlo}+\textsc{Pythia8}~\cite{vbfnlo,pythia}
for the hard $Hjj$ production process with the scale choice
$\mu_F^2=\mu_R^2=m_H^2+p_{T,j1}^2+p_{T,j2}^2$.  For example the slight
broadening of the low-$p_T$ peaks compared to
Fig.~\ref{fig:partonic_log} is due to parton shower effects and this
scale choice.  The full simulation confirms that the $Hj$ process has
a larger total rate than the $Hjj$ process, but this additional $Hj$
rate is concentrated at small transverse momenta and does not carry
information on the Higgs--gluon coupling. For $p_{T,H}>300$~GeV the
parton shower is expected to underestimate additional jet radiation
off the $Hj$ process and cannot be expected to reflect the top mass
logarithms; hence, the $Hjj$ process gives a larger relevant number of
events to probe the Higgs--gluon vertex.  This is universally true for
all three contributions defined in Eq.\eqref{eq:distribution}.

\section{Signal--background analyses}

Following the results in the last section the key question becomes how
much, in addition to the increase in the number of relevant signal
events, the background rejection benefits from the additional jet in
the hard process.  As simple examples we consider the two most
promising Higgs decay channels, $H \to WW$ and $H \to \tau\tau$ in the
fully leptonic decay modes at the LHC with $\sqrt{S}=13$~TeV.\bigskip

The signal events are generated with \textsc{Mcfm}~\cite{mcfm} for the
$Hj$ process and with \textsc{Vbfnlo}~\cite{vbfnlo} for the $Hjj$
process, respectively. They are showered with
\textsc{Pythia8}~\cite{pythia}. Both generators provide results for
finite top mass, $\kappa_{t,g}=(1,0)$, as well as the pure dimension-6
scenario $\kappa_{t,g}=(0,1)$.  To probe the whole $\kappa_t$ vs
$\kappa_g$ range we expand both implementations including the complete
interference structure given in Eq.\eqref{eq:distribution}.  Because
there are no next--to--leading order (NLO) computations available for
either of the two channels with full top mass dependence, we scale our
total cross sections to the corresponding NLO rates in the heavy top
limit. For a consistent scale choice we apply a flat correction of
$K_{Hj}\sim1.4$~\cite{mcfm} and $K_{Hjj}\sim1.6$~\cite{gosam}. In
particular for the 2 jet case, it is known that distributions are
reproduced at the 10\% level even for highly boosted Higgs bosons when
the full top mass dependence is included at LO \cite{robert}.

The $t\bar{t}$+jets and $WW$+jets background are generated with the
\textsc{PowhegBox}~\cite{powheg}, showered with a vetoed
\textsc{Pythia8} shower~\cite{pythia}.  We also include the $Z$+jets
background from \textsc{Sherpa}+\textsc{BlackHat}~\cite{sherpa_z}
merged at next--to--leading order with up to three hard jets. In all
background processes we enforce top, $W$, and $Z$ decays to
charged leptons, \ie muons, electrons or taus. Jets are
defined using the anti-$k_T$ algorithm implemented in
\textsc{Fastjet}~\cite{fastjet} with $R=0.5$ and
\begin{alignat}{5}
p_{T,j}>40~\gev 
\qquad \text{and} \qquad 
|y_j|<4.5 \;.
\label{eq:cut_jtau}
\end{alignat} 
If explicitly shown, the one or two recoil jets are defined as the
hardest jets fulfilling this requirement. Throughout we smear the
missing energy vector using a gaussian. For the leptons we require two
isolated opposite sign leptons with
\begin{alignat}{5}
p_{T,\ell}>20~\gev
\qquad \text{and} \qquad 
|y_\ell|<2.5 \;,
\label{eq:cut_jtau}
\end{alignat} 
where the isolation criterion is a hadronic energy deposition
$E_{T,\text{had}} < E_{T,\ell}/10$ within a cone of size $R=0.2$.  To
suppress the top background we require zero $b$-tags with a flat
tagging efficiency of $70\%$ and a mistag rate of $2\%$. Our
simulation of the top pair background should be taken with a grain of
salt, because there are many ways of further suppressing this
background based on the underlying jet structure~\cite{fwm}. Note that
the focus of this signal and background analysis is not to estimate a
realistic target for the measurement of $\kappa_t$ and $\kappa_g$, but
to see how the $Hjj$ process compares with the $Hj$
process~\cite{sanz,azatov,andi,schlaffer_spannowsky}.

\subsection*{$H\to WW$ decays}
\label{sec:w}

As the first signature, we show how we can probe the structure of the
Higgs--gluon coupling in $Hjj$ production based on leptonic $H \to WW$
decays. To estimate the additional benefit of including the second jet
we compare the signal--to--background ratios $S/B$ for Higgs
production with one and two hard jets. For the $WW$ decay channel the
main backgrounds are $WW$+jets and $t\bar{t}$+jets production. We
start with the basic cuts shown in the first lines of
Tab.~\ref{tab:cuts1}.

\begin{table}[b!]
\begin{tabular}{l || c | c | c || c | c | c  }
  \multicolumn{1}{c||}{} &
  \multicolumn{3}{c||}{$Hj \to (WW)j$ inclusive}&
  \multicolumn{3}{c}{$Hjj \to (WW)jj$ inclusive}
    \\
  \hline
 \multirow{1}{*}{cuts} &  
 \multicolumn{1}{c|}{$H$+jets} & 
 \multicolumn{1}{c|}{$WW$+jets}  & 
 \multicolumn{1}{c||}{$t\bar{t}$+jets}&
 \multicolumn{1}{c|}{$H$+jets} & 
 \multicolumn{1}{c|}{$WW$+jets}  & 
  \multicolumn{1}{c}{$t\bar{t}$+jets} \\ 
  \hline
{$p_{T,j}>40$~\gev, $|y_j|<4.5$}
&\multirow{2}{*}{35.5 } & \multirow{2}{*}{524} &\multirow{2}{*}{14770 } 
&\multirow{2}{*}{ 17.3} & \multirow{2}{*}{90.7} &\multirow{2}{*}{7633}\\  
$p_{T,\ell}>20$~\gev, $|y_\ell|<2.5$
 &  & & & &  & \\  
$N_b=0$                
& 33.3 & 515 & 4920
& 15.2 & 87.4 & 1690  \\ 
$m_{\ell \ell}\in [10,60]~\gev$               
& 28.3 & 106 & 1060  
& 13.0 & 17.2 & 351  \\ 
$\slashchar{E}_T>45~\gev$               
& 21.4 & 92.9 & 930  
& 10.6 & 15.9 & 309   \\
$\Delta \phi_{\ell \ell}<0.8$               
& 14.3 & 49.8 & 479  
& 8.14 & 10.3 & 162   \\
$m_T<125~\gev$               
& 14.2 & 26.6 & 220  
& 8.09 & 6.14 & 76.2   \\
$p_{T,H}>300~\gev$               
& 0.59 & 2.73 & 5.18  
& 1.06 & 1.39 & 3.28  \\ \hline
$\Delta \phi_{jj}<1.8$               
&&&
& 0.87 & 1.05 & 1.33   \\
$p_{T,j1}/p_{T,j2}<2.5$               
&&&
& 0.57 & 0.53 & 0.53  
\end{tabular} 
\caption{Cut flow for $H$+jets, $WW$+jets and $t\bar{t}$+jets. All
  rates are given in fb.}
\label{tab:cuts1}
\end{table}

Aside from the missing weak boson fusion
characteristics they are similar to the known analysis techniques for
Higgs production in association with two jets. Obviously, we do not
apply a stiff $m_{jj}$ cut to reduce QCD backgrounds as well as gluon
fusion Higgs production. The transverse mass of the $WW$ system is
defined as
\begin{alignat}{5}
m_T^2 = (E_T^{\ell \ell}+\slashchar{E}_T)^2
      -|\vec{p}_T^{\,\ell\ell}+\slashchar{\vec{E}}_T|^2
\qquad \text{with} \qquad  E_T^{\ell \ell}=\sqrt{|\vec{p}_T^{\, \ell \ell}|^2+m_{\ell \ell}^2}\;.
\label{eq:mT}
\end{alignat} 
The $p_{T,H}$ cut extracts events which are sensitive to the
logarithmic dependence on the top mass. The numbers shown for the $Hj$
process are in good agreement with the findings of
Ref.~\cite{schlaffer_spannowsky}.  As expected from
Fig.~\ref{fig:mother_distrib} the number of signal events in the $Hjj$
process exceeds the corresponding number in the $Hj$ channel by a
factor of two. Moreover, in particular the $WW$+jets background is
reduced by the required second hard jet.\bigskip

\begin{figure}[!t]
 \includegraphics[width=0.35\textwidth]{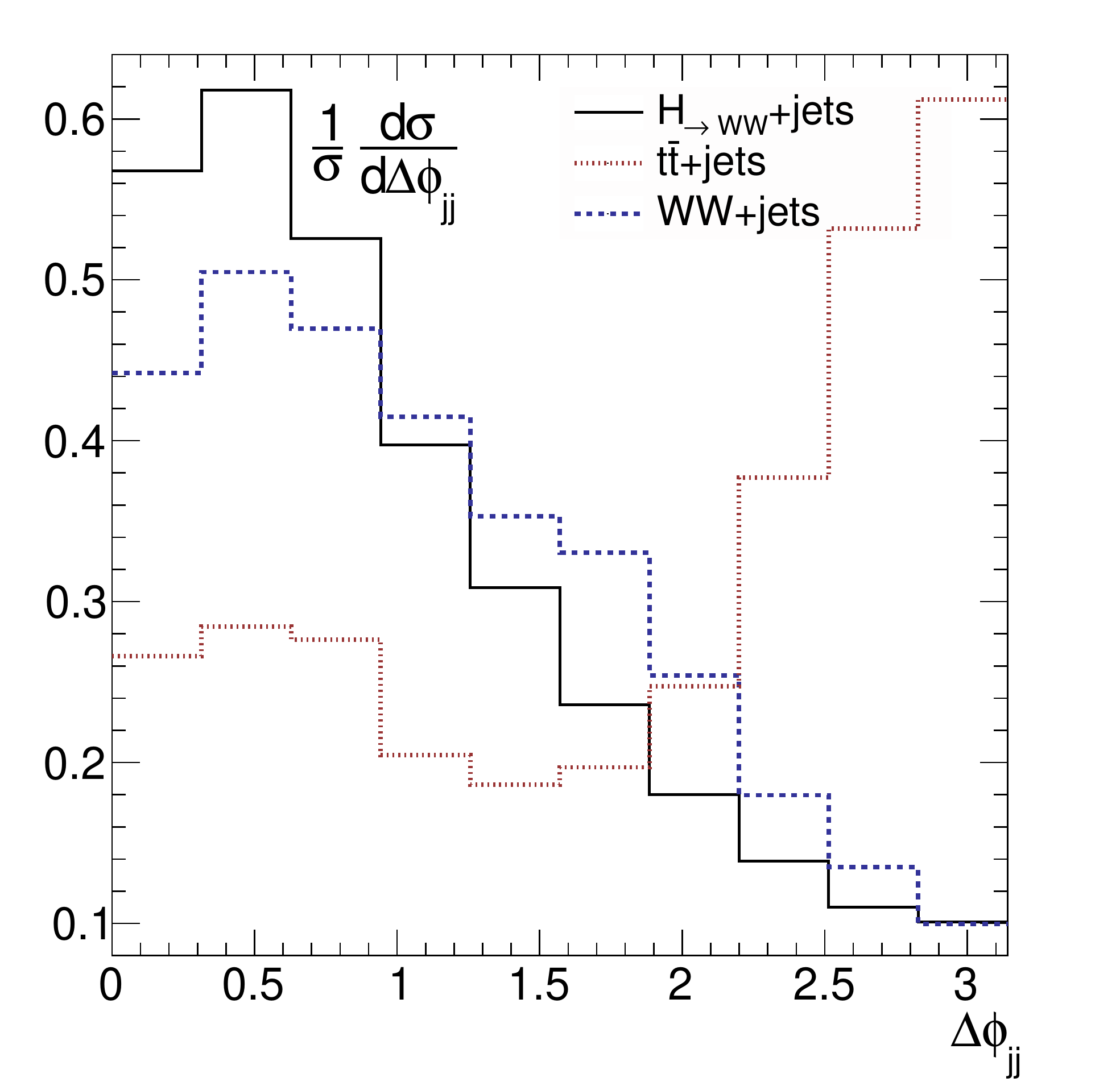}
   \hspace*{0.1\textwidth}
  \includegraphics[width=0.35\textwidth]{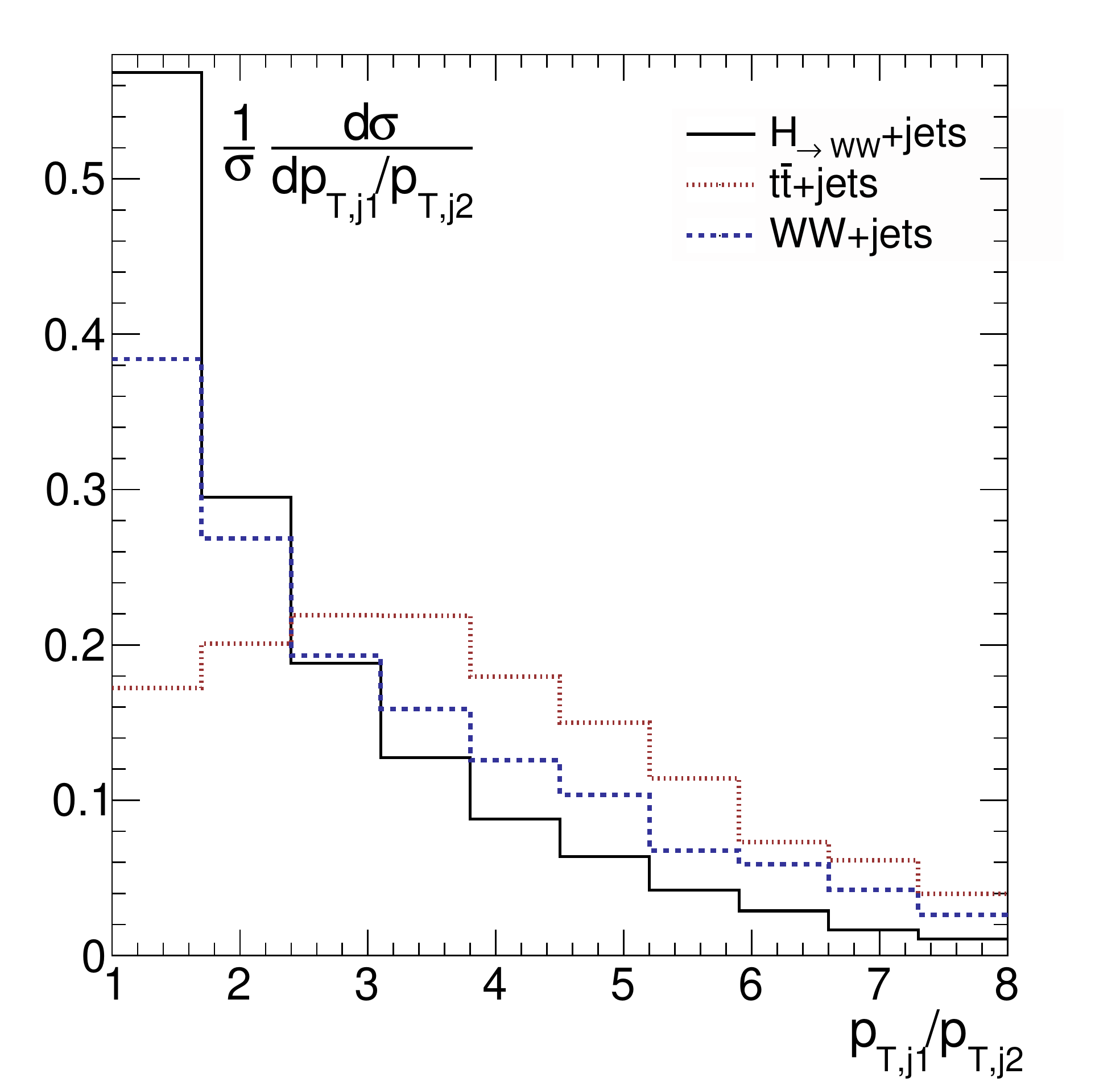}
\caption{Normalized $\Delta \phi_{jj}$ (left) and $p_{T,j1}/p_{T,j2}$
  (right) distributions for the $H\rightarrow WW$ signal and the
  dominant backgrounds. All universal cuts listed in
  Tab~\ref{tab:cuts1} are already applied.}
\label{fig:distri_ww} 
\end{figure}

In addition, we can use the second jet to define additional
observables which can in turn be used to suppress backgrounds. Two
choices, namely the azimuthal angle between the tagging
jets~\cite{phi_jj} and the ratio of transverse momenta of the two
jets, are shown in Fig.~\ref{fig:distri_ww}. It is interesting to
notice that the usual application of the azimuthal angle between the
tagging jets relies on the forward jet kinematics, while in this
analysis the tagging jets are hard and relatively central. First, we
see that the boosted Higgs configuration forces the two recoil jets
for the Higgs signal and the $WW$ background to move close to each
other in the azimuthal plane. In addition, two jets recoiling against
one Higgs boson prefers more balanced jet momenta than the recoil
against two independently produced $W$ bosons. This again supports our
earlier conclusion that the underlying hard process indeed includes
two hard jets.  Cutting on both jet--jet correlations we can reduce
the $WW$ background to the $Hjj$ signal to roughly a fifth of the
corresponding $Hj$ background, for similar signal rates in the boosted
phase space region.

\subsection*{$H\to \tau\tau$ decays}
\label{sec:tau}

\begin{table}[b!]
\begin{tabular}{l || c | c | c | c || c | c | c | c  }
  \multicolumn{1}{c||}{} &
  \multicolumn{4}{c||}{$Hj \to (\tau\tau)j$ inclusive}&
  \multicolumn{4}{c}{$Hjj \to (\tau\tau)jj$ inclusive}
    \\
  \hline
 \multirow{1}{*}{cuts} &  
 \multicolumn{1}{c|}{$H$+jets} & 
 \multicolumn{1}{c|}{$Z/\gamma^*$+jets}  & 
 \multicolumn{1}{c|}{$WW$+jets}&
 \multicolumn{1}{c||}{$t\bar{t}$+jets}&
 \multicolumn{1}{c|}{$H$+jets} & 
 \multicolumn{1}{c|}{$Z/\gamma^*$+jets}  & 
 \multicolumn{1}{c|}{$WW$+jets}&
  \multicolumn{1}{c}{$t\bar{t}$+jets} \\ 
  \hline
{$p_{T,j}>40$~\gev, $|y_j|<4.5$}
&\multirow{2}{*}{9.82 } & \multirow{2}{*}{162303} &\multirow{2}{*}{524}&\multirow{2}{*}{14770 } 
&\multirow{2}{*}{ 5.10} & \multirow{2}{*}{27670} &\multirow{2}{*}{90.7} &\multirow{2}{*}{7633}\\  
$p_{T,\ell}>20$~\gev, $|y_\ell|<2.5$
 &  & & & & & &  & \\  
$N_b=0$                
& 9.21 & 148221 & 515 & 4920
& 4.50 &  23218  & 87.4 & 1690  \\ 
{$m_{\ell \ell} \in [10,60]~\gev$}
&\multirow{2}{*}{6.59 } & \multirow{2}{*}{10466} &\multirow{2}{*}{179} &\multirow{2}{*}{1616 } 
&\multirow{2}{*}{3.41} & \multirow{2}{*}{1832} &\multirow{2}{*}{28.3}&\multirow{2}{*}{541}\\  
$m_{\ell \ell'} \in [10,100]~\gev$               
 &  & & & &  &&& \\  
$\slashchar{E}_T>45~\gev$               
& 6.24 & 38.1 &166 & 1616  
& 3.31 &  0.65& 27.0 & 541   \\
$|m_{\tau\tau}-m_H|<20~\gev$               
& 5.88 & 2.84 & 6.28 & 45.9  
& 3.10 & 0.11 & 1.18 & 16.0   \\
$p_{T,H}>300~\gev$               
& 0.23 & 0.013  & 0.40 & 0.87  
& 0.41 & 0.004  & 0.20 & 0.56
  \\ \hline
$\Delta \phi_{jj}<1.8$               
&&&&
& 0.33 & 0  & 0.15& 0.22   \\
$p_{T,j1}/p_{T,j2}<2.5$               
&&&&
& 0.22 & 0 &0.076 & 0.086 
\end{tabular} 
\caption{Cut flow for $H$+jets, $Z/\gamma^*$+jets, $WW$+jets and
  $t\bar{t}$+jets. All rates are given in fb.}
\label{tab:cuts2}
\end{table}

\begin{figure}[!t]
 \includegraphics[width=0.35\textwidth]{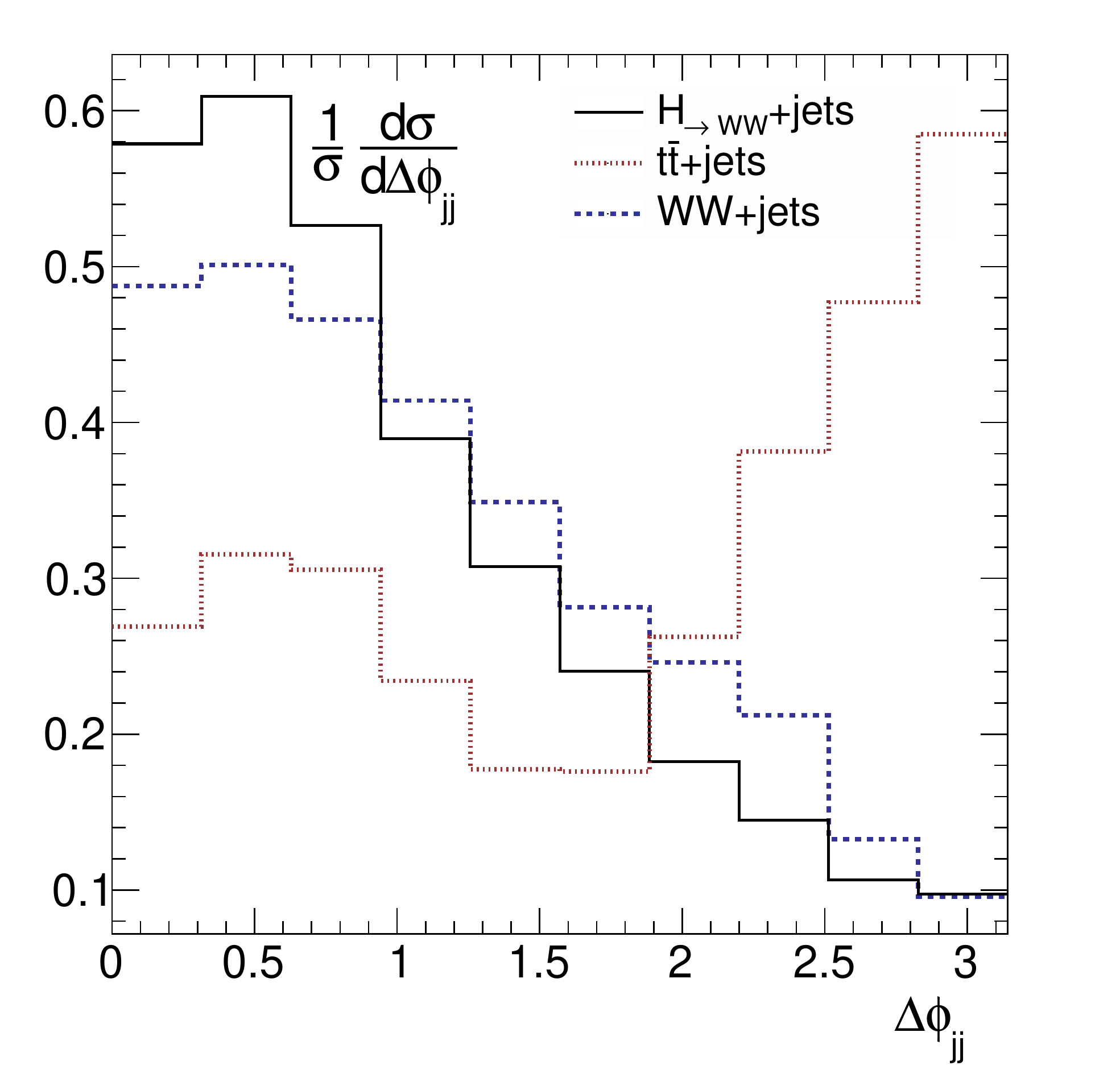}
   \hspace*{0.1\textwidth}
 \includegraphics[width=0.35\textwidth]{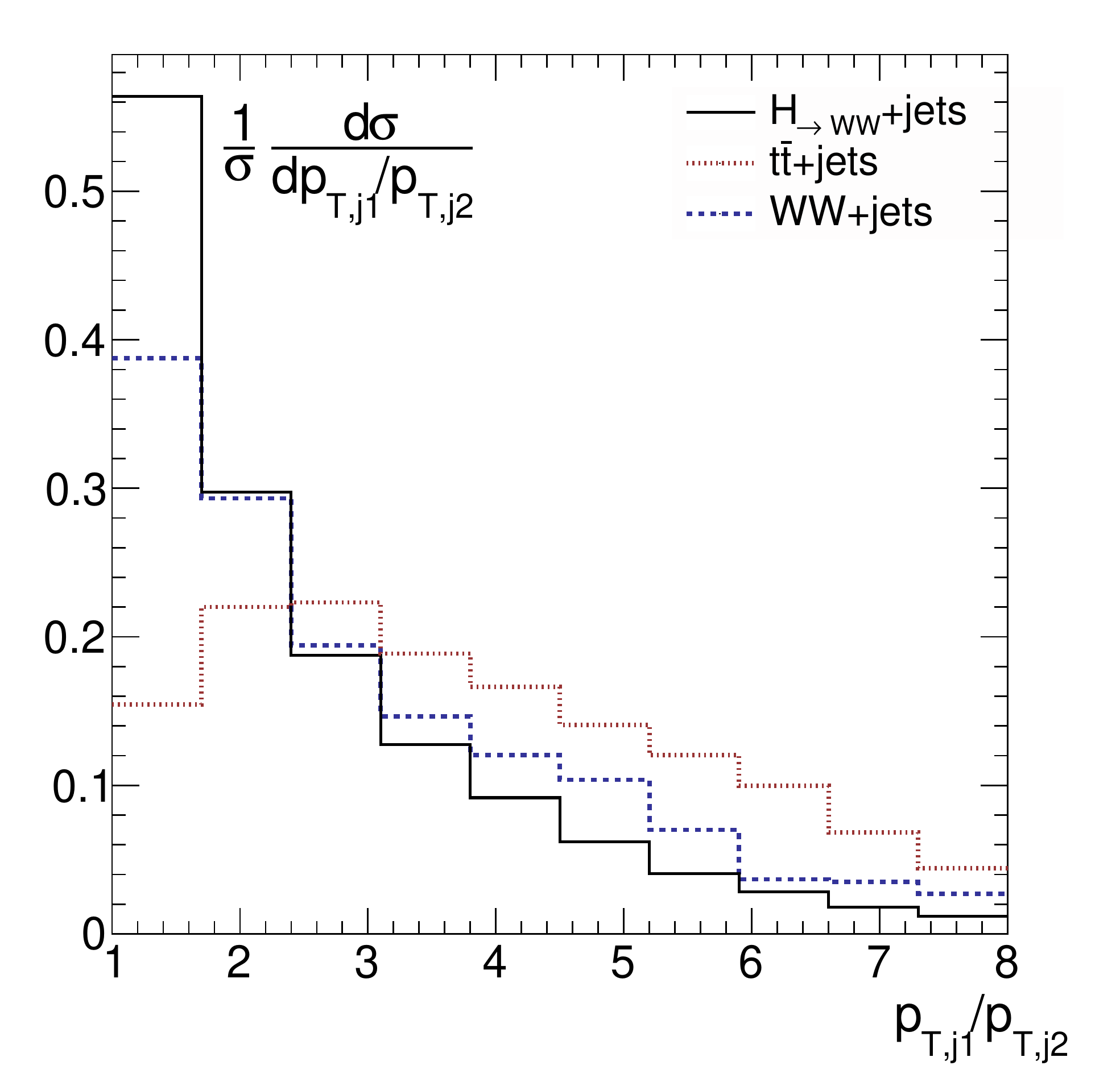}
\caption{Normalized $\Delta \phi_{jj}$ (left) and $p_{T,j1}/p_{T,j2}$
  (right) distributions for the $H\rightarrow \tau\tau$ signal and the
  dominant backgrounds. All universal cuts listed in
  Tab~\ref{tab:cuts2} are already applied.}
\label{fig:distri_tautau} 
\end{figure}

As an alternative decay signature we also study $Hjj$ production with
a purely leptonic $H \to \tau \tau$ decay.  Because the leptonic $WW$
and $\tau\tau$ decay channels have a similar detector signature and
main backgrounds are $t\bar{t}$+jets and $WW$+jets we stick to a
similar initial analysis strategy, now shown in
Tab.~\ref{tab:cuts2}. Instead of the transverse mass cut we compute
$m_{\tau \tau}$ in the collinear approximation~\cite{keith},
\begin{alignat}{5}
m_{\tau\tau}=\frac{m_\text{vis}}{ \sqrt{x_1 x_2}} 
\qquad \text{with} \qquad 
x_{1,2}=\frac{p_{\text{vis}1,2}}{p_{\text{vis}1,2}+p_{\text{miss}1,2}} \;,
\label{eq:cut_jtau}
\end{alignat} 
where $m_\text{vis}$ and $p_\text{vis}$ are the invariant mass and
total momentum of the visible tau decay products. The variable
$p_\text{miss}$ is the neutrino momentum reconstructed in the
collinear approximation. Using this approximation we require
\begin{alignat}{5}
|m_{\tau\tau}-m_H|<20~\gev 
\qquad \text{with} \qquad 
x_{1,2} \in [0.1,1] \; .
\label{eq:cut_mtautau}
\end{alignat} 
This large mass window should include the vast majority of signal
events while we will see that it is still sufficient to control the
backgrounds. By imposing
\begin{alignat}{5}
p_{T,H} \sim p_{T,\ell_1} + p_{T,\ell_2} + \slashchar{p}_T > 300~\gev 
\end{alignat}
we ensure perfect kinematical conditions to apply the collinear
approximation.  Similar to the $WW$ channel we then use the second jet
to further suppress the backgrounds, see
Fig.~\ref{fig:distri_tautau}. As for the $WW$ case we see that for
similar event numbers in the top--mass--sensitive region the
backgrounds in the $Hjj$ analysis are something like a factor 1/5
smaller than for the $Hj$ case.\bigskip

\begin{figure}[!b]
 \includegraphics[width=0.325\textwidth]{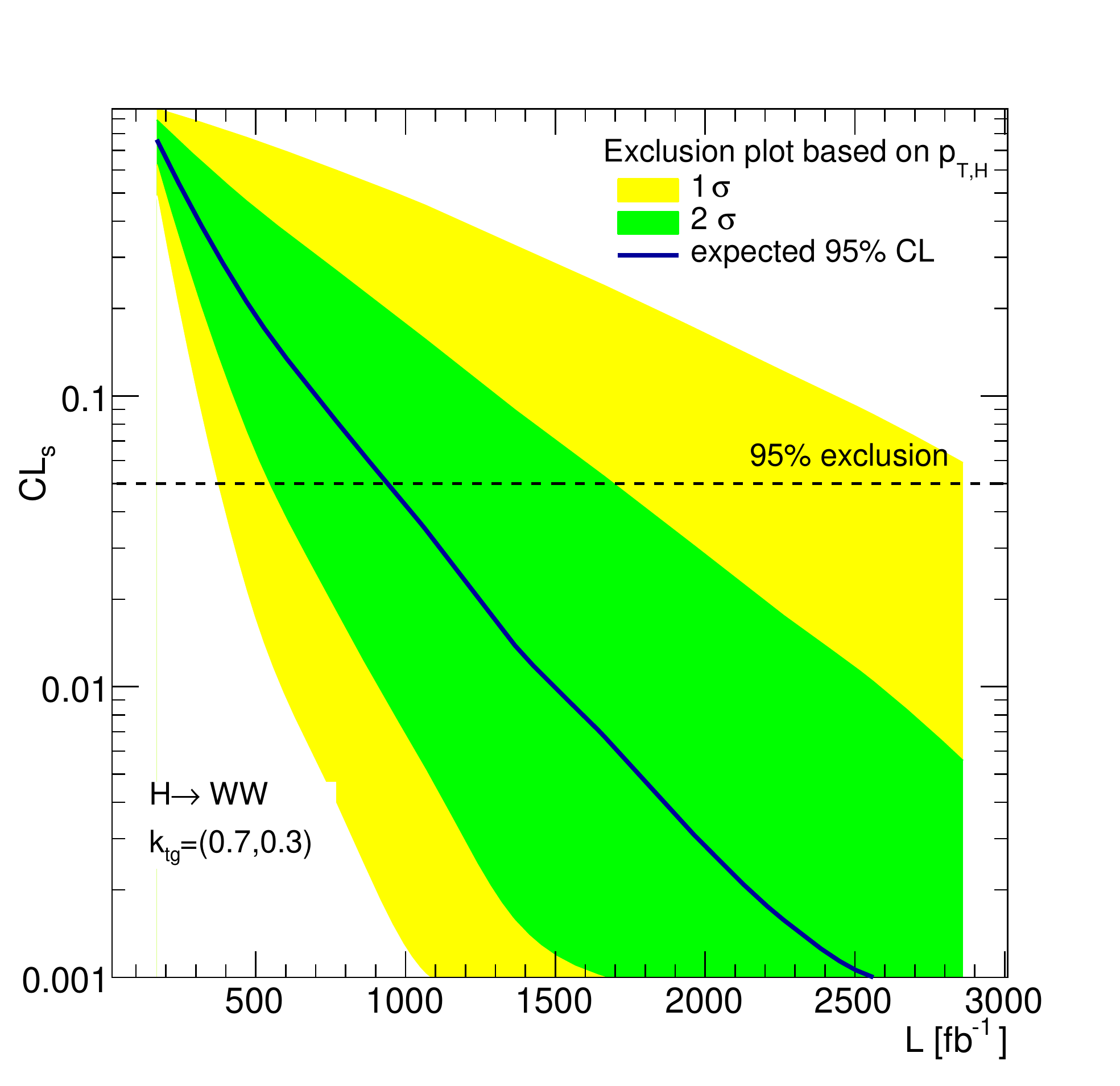}
 \includegraphics[width=0.325\textwidth]{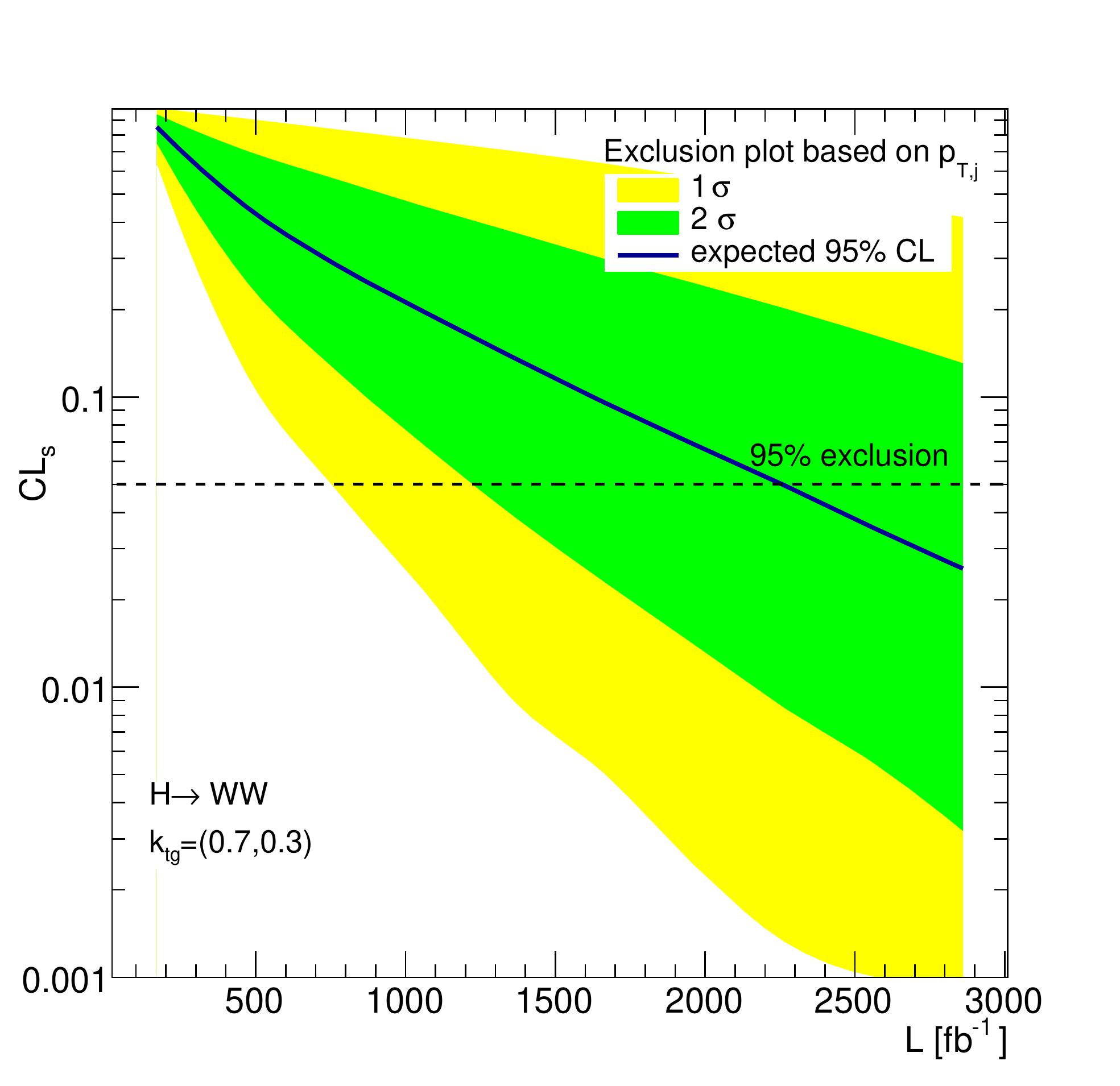}
 \includegraphics[width=0.325\textwidth]{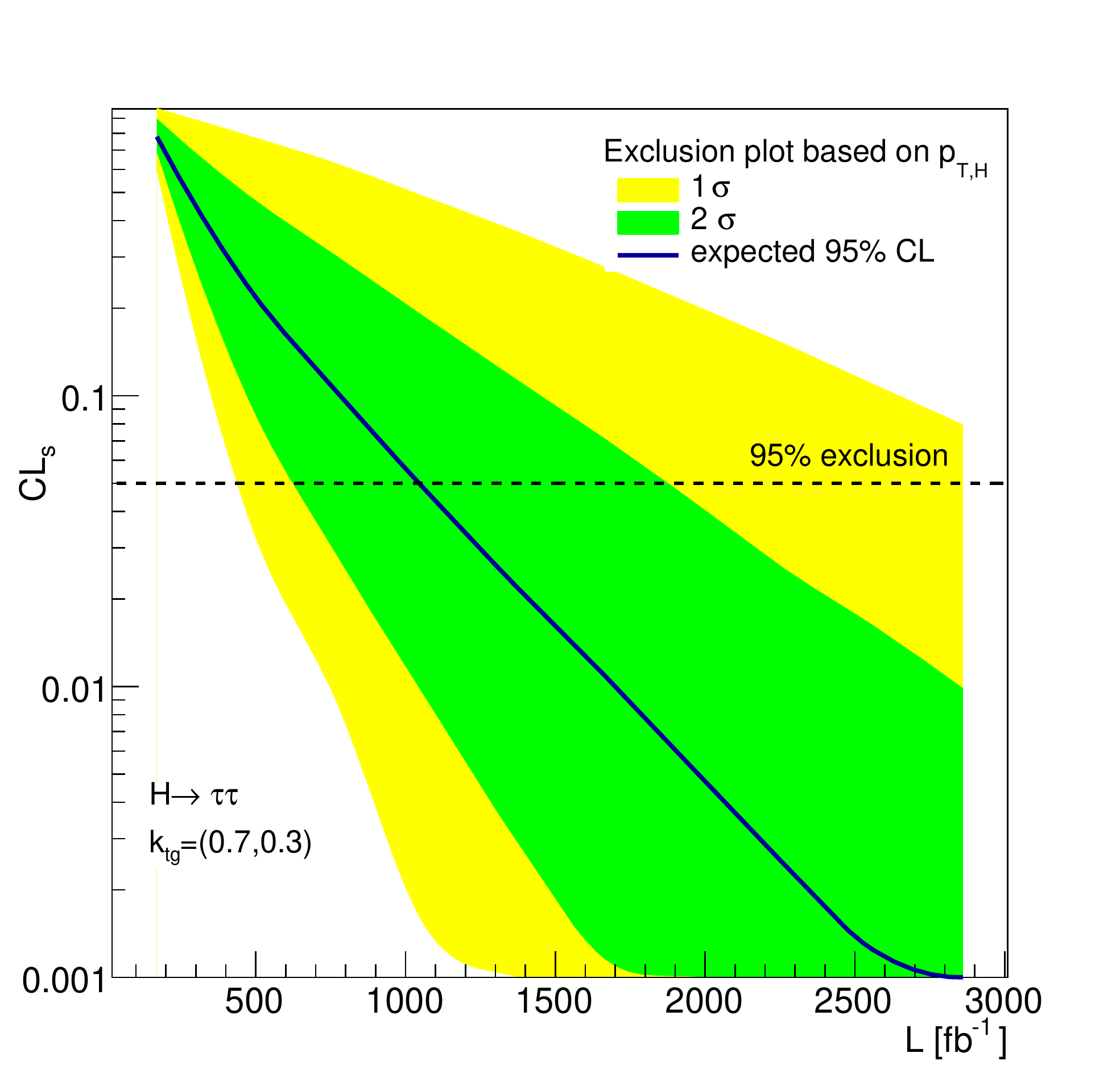}
\caption{Confidence level for separating the BSM hypotheses
  $\kappa_{t,g}=(0.7,0.3)$ from the Standard Model. We show results
  for $H\rightarrow WW$ decays based on the transverse momentum of the
  Higgs (left) and the hardest jet (center). For the $H\rightarrow
  \tau \tau$ decays (right) we limit ourselves to the more promising case of
  the Higgs transverse momentum.}
\label{fig:CLs} 
\end{figure}

Combining the different $p_{T,H}$ bins into a shape analysis allows us
to extract information on the parameters $\kappa_t$ and $\kappa_g$
introduced in Eq.\eqref{eq:lagrangian}.  To estimate the power of the
$Hjj$ analysis we evaluate the $p_{T,H}$ distribution using the
$\text{CL}_s$ method. The Standard Model $\kappa_{tg}=(1,0)$ defines
the null hypothesis, to be compared with the BSM parameter point
$\kappa_{tg}=(0.7,0.3)$.  For the results shown in Fig.~\ref{fig:CLs}
we assume a NLO scale uncertainty of
$\mathcal{O}(20\%)$~\cite{vbfnlo}. We also show results for the
leading $p_{T,j}$ distribution, indicating that the Higgs transverse
momentum is the best--suited single observable for the $Hjj$
analysis. Unlike for the $Hj$ analysis we find that the leptonic $WW$
and $\tau \tau$ decays are similarly promising.\bigskip 

As indicated by Fig.~\ref{fig:CLs}, excluding small deviations of the
Higgs--top and Higgs--gluon from their Standard Model values couplings
remains a challenge for the upcoming LHC runs. To accumulate the best
sensitivity possible it will be necessary to combine all available
channels. However, cleanly separating the leptonic $H\to \tau\tau$ and
$H\to WW$ decays in $Hj$ production is kinematically very
difficult~\cite{schlaffer_spannowsky}. In this study we now find that
$Hjj$ production with a decay $H \to WW$ is almost as sensitive as
the corresponding $\tau \tau$ decay channel, so with full control over
the additional one or two jets a combination of the two decay channels
seems possible.

\section{Conclusions}
\label{sec:conclusions}

We have shown that the extraction of the top mass dependence in the
effective Higgs--gluon coupling at the LHC benefits from a second jet,
\ie a hard process consisting of the Higgs plus two jets. As two
robust example signatures we consider purely leptonic Higgs decays to
$W$ bosons and $\tau$ leptons.  Higgs production with two hard jets
should not be considered a correction to Higgs production plus one jet
in the boosted regime, because in the corresponding analysis we find:
\begin{enumerate}
\item the divergence structure of the $Hjj$ process is given by a
  similar logarithm as the $Hj$ case; numerically, the VBF topology
  with two hard jets radiated off the initial state partons dominates
  the top mass dependence at large transverse momenta.
\item adding a second hard jet moves a large fraction of signal events
  from top--mass--insensitive phase space regions to
  top--mass--sensitive configurations. For large transverse momenta of
  the Higgs boson the $Hjj$ production process even contributes more
  signal events than the $Hj$ process.
\item a second fully correlated jet described by the hard matrix
  element can be used to reduce the backgrounds by roughly a factor
  1/5 for a similar number of signal events, compared to the same
  analysis with only one hard jet.
\item both, the $H \to WW$ and $H \to \tau\tau$ signatures appear
  feasible when combined with the $Hjj$ production process.
\end{enumerate}
Given the statistical limitation of this detailed study of the
Higgs--gluon coupling and its underlying loop structure the $Hjj$
channel should be a very useful additional handle.  Obviously, a fully
merged analysis of the $Hj$ and $Hjj$ channels including the complete
Higgs--gluon coupling structure will combine the two available
channels for example in the $p_{T,H}$ distribution.

\acknowledgments 

We would like to thank Marek Sch\"onherr for his kind help with all
kinds of QCD issues and Michael Spira for valuable discussions. CE is
supported by the Institute for Particle Physics Phenomenology
Associateship program.



\begin{thebibliography}{99}

\bibitem{higgs}
 P.~W.~Higgs,
  Phys.\ Lett.\  {\bf 12}, 132 (1964);
 P.~W.~Higgs,
  Phys.\ Rev.\ Lett.\  {\bf 13}, 508 (1964);
  P.~W.~Higgs,
  and Phys.\ Rev.\  {\bf 145}, 1156 (1964);
 F.~Englert and R.~Brout,
  Phys.\ Rev.\ Lett.\  {\bf 13}, 321 (1964);
  G.S.~Guralnik, C.R.~Hagen and T.W.~Kibble,
  Phys.\ Rev.\ Lett.\  {\bf 13}, 585 (1964).

\bibitem{discovery}
 G.~Aad {\it et al.} [ATLAS Collaboration],
 Phys.\ Lett.\ B {\bf 716}, 1 (2012),
 S.~Chatrchyan {\it et al.}  [CMS Collaboration],
 Phys.\ Lett.\ B {\bf 716}, 30 (2012).

\bibitem{sfitter} 
  M.~Klute, R.~Lafaye, T.~Plehn, M.~Rauch and D.~Zerwas,
  Phys.\ Rev.\ Lett.\  {\bf 109}, 101801 (2012);
  D.~Lopez-Val, T.~Plehn and M.~Rauch,
  JHEP {\bf 1310}, 134 (2013).

\bibitem{couplings_ex}
  The ATLAS collaboration, ATLAS-CONF-2012-170.
  The CMS collaboration, CMS-PAS-HIG-12-045.

\bibitem{couplings_th}
  A.~Azatov, R.~Contino and J.~Galloway,
  JHEP {\bf 1204}, 127 (2012);
  P.~P.~Giardino, K.~Kannike, M.~Raidal and A.~Strumia,
  Phys.\ Lett.\ B {\bf 718}, 469 (2012);
  J.~Ellis and T.~You,
  JHEP {\bf 1209}, 123 (2012);
  J.~R.~Espinosa, C.~Grojean, M.~M\"uhlleitner and M.~Trott,
  JHEP {\bf 1205}, 097 (2012), 
  JHEP {\bf 1209}, 126 (2012) 126, 
  and
  JHEP {\bf 1212}, 045 (2012);
  A.~Djouadi and G.~Moreau,
  arXiv:1303.6591 [hep-ph];
  J.~Ellis and T.~You,
  JHEP {\bf 1306}, 103 (2013).

\bibitem{higgs_review}
  For recent discussions of dimension-6 operators see \textsl{e.g.}
  A.~Azatov and J.~Galloway,
  Int.\ J.\ Mod.\ Phys.\ A {\bf 28} (2013) 1330004;
  I.~Brivio {\it et al.},
  JHEP {\bf 1403} (2014) 024;
  R.~Contino, M.~Ghezzi, C.~Grojean, M.~Muhlleitner and M.~Spira,
  JHEP {\bf 1307} (2013) 035;
  J.~Elias-Miro, J.~R.~Espinosa, E.~Masso and A.~Pomarol,
  JHEP {\bf 1311} (2013) 066;
  C.~Englert, A.~Freitas, M.~M\"uhlleitner, T.~Plehn, M.~Rauch, M.~Spira and K.~Walz,
  arXiv:1403.7191 [hep-ph];
  J.~Ellis, V.~Sanz and T.~You,
  arXiv:1404.3667 [hep-ph].

\bibitem{tth}
  A.~Belyaev and L.~Reina,
at the LHC,''
  JHEP {\bf 0208}, 041 (2002);
   E.~Gross and L.~Zivkovic,
  Eur.\ Phys.\ J.\ {\bf 59}, 731 (2009);
  T.~Plehn, G.~P.~Salam and M.~Spannowsky,
  Phys.\ Rev.\ Lett.\  {\bf 104} (2010) 111801;
  C.~Boddy, S.~Farrington and C.~Hays,
  Phys.\ Rev.\ D {\bf 86}, 073009 (2012);
  P.~Artoisenet, P.~de Aquino, F.~Maltoni and O.~Mattelaer,
  Phys.\ Rev.\ Lett.\  {\bf 111} (2013) 9,  091802;
  P.~Agrawal, S.~Bandyopadhyay and S.~P.~Das,
  arXiv:1308.6511 [hep-ph].
  M.~R.~Buckley, T.~Plehn, T.~Schell and M.~Takeuchi,
  JHEP {\bf 1402}, 130 (2014).

\bibitem{thj}
  M.~Farina, C.~Grojean, F.~Maltoni, E.~Salvioni and A.~Thamm,
  JHEP {\bf 1305}, 022 (2013);
  S.~Biswas, E.~Gabrielli, F.~Margaroli and B.~Mele,
  JHEP {\bf 07}, 073 (2013);
  J.~Ellis, D.~S.~Hwang, K.~Sakurai and M.~Takeuchi,
  JHEP {\bf 1404} (2014) 004;
  C.~Englert and E.~Re,
  Phys.\ Rev.\ D {\bf 89}, 073020 (2014).

\bibitem{th} 
 W.~J.~Stirling and D.~J.~Summers,
  Phys.\ Lett.\ B {\bf 283}, 411 (1992).
 F.~Maltoni, D.~L.~Rainwater and S.~Willenbrock,
  Phys.\ Rev.\ D {\bf 66} (2002) 034022.

\bibitem{sm_only}
 see \textsl{e.g.}
 M.~Shaposhnikov and C.~Wetterich,
  Phys.\ Lett.\ B {\bf 683}, 196 (2010);
 M.~Holthausen, K.~S.~Lim and M.~Lindner,
  JHEP {\bf 1202}, 037 (2012);
 A.~Hebecker, A.~K.~Knochel and T.~Weigand,
  Nucl.\ Phys.\ B {\bf 874}, 1 (2013);
 D.~Buttazzo {\it et al}
  JHEP {\bf 1312}, 089 (2013).

\bibitem{bsm_review}
  D.~E.~Morrissey, T.~Plehn and T.~M.~P.~Tait,
  Phys.\ Rept.\  {\bf 515}, 1 (2012).

\bibitem{low_energy}
  J.~R.~Ellis, M.~K.~Gaillard and D.~V.~Nanopoulos,
  Nucl.\ Phys.\ B {\bf 106}, 292 (1976);
  M.~A.~Shifman, A.~I.~Vainshtein, M.~B.~Voloshin and V.~I.~Zakharov,
  Sov.\ J.\ Nucl.\ Phys.\  {\bf 30}, 711 (1979)
  [Yad.\ Fiz.\  {\bf 30}, 1368 (1979)];
  B.~A.~Kniehl and M.~Spira,
  Z.\ Phys.\ C {\bf 69}, 77 (1995).

\bibitem{lecture}
  for a pedagogical introduction see \eg 
 T.~Plehn,
  Lect.\ Notes Phys.\  {\bf 844}, 1 (2012). 

\bibitem{spirix_nlo}
  D.~Graudenz, M.~Spira and P.~M.~Zerwas,
  Phys.\ Rev.\ Lett.\  {\bf 70}, 1372 (1993);
  M.~Spira, A.~Djouadi, D.~Graudenz and P.~M.~Zerwas,
  Nucl.\ Phys.\ B {\bf 453}, 17 (1995);
  M.~Kr\"amer, E.~Laenen and M.~Spira,
  Nucl.\ Phys.\ B {\bf 511}, 523 (1998);
  S.~Marzani, R.~D.~Ball, V.~Del Duca, S.~Forte and A.~Vicini,
  Nucl.\ Phys.\ B {\bf 800}, 127 (2008),
  A.~Pak, M.~Rogal and M.~Steinhauser,
  JHEP {\bf 1002}, 025 (2010).

\bibitem{robert}
  R.~V.~Harlander, T.~Neumann, K.~J.~Ozeren and M.~Wiesemann,
  JHEP {\bf 1208}, 139 (2012).


\bibitem{spirix_review}
  M.~Spira,
  Fortsch.\ Phys.\  {\bf 46}, 203 (1998).

\bibitem{higgs_pair}
  T.~Plehn, M.~Spira and P.~M.~Zerwas,
  Nucl.\ Phys.\ B {\bf 479}, 46 (1996)
  [Erratum-ibid.\ B {\bf 531}, 655 (1998)];
  U.~Baur, T.~Plehn and D.~L.~Rainwater,
  Phys.\ Rev.\ Lett.\  {\bf 89}, 151801 (2002);
  M.~J.~Dolan, C.~Englert and M.~Spannowsky,
  JHEP {\bf 1210} (2012) 112.
  J.~Grigo, J.~Hoff, K.~Melnikov and M.~Steinhauser,
  Nucl.\ Phys.\ B {\bf 875}, 1 (2013);
  Phys.\ Rev.\ D {\bf 89}, 013012 (2014).

\bibitem{light_stop}
  D.~S.~M.~Alves, M.~R.~Buckley, P.~J.~Fox, J.~D.~Lykken and C.~-T.~Yu,
  Phys.\ Rev.\ D {\bf 87}, no. 3, 035016 (2013);
  S.~Bornhauser, M.~Drees, S.~Grab and J.~S.~Kim,
  Phys.\ Rev.\ D {\bf 83}, 035008 (2011)
  N.~Desai and B.~Mukhopadhyaya,
  JHEP {\bf 1205}, 057 (2012);
  Z.~Han, A.~Katz, D.~Krohn and M.~Reece,
  JHEP {\bf 1208}, 083 (2012);
  G.~Belanger, R.~M.~Godbole, L.~Hartgring and I.~Niessen,
  JHEP {\bf 1305}, 167 (2013);
  X.~-Q.~Li, Z.~-G.~Si, K.~Wang, L.~Wang, L.~Zhang and G.~Zhu,
  Phys.\ Rev.\ D {\bf 89}, 077703 (2014).

\bibitem{keith}
  R.~K.~Ellis, I.~Hinchliffe, M.~Soldate and J.~J.~van der Bij,
  Nucl.\ Phys.\ B {\bf 297}, 221 (1988).

\bibitem{uli}
  U.~Baur and E.~W.~N.~Glover,
  Nucl.\ Phys.\ B {\bf 339}, 38 (1990).

\bibitem{schlaffer_spannowsky}
  M.~Schlaffer, M.~Spannowsky, M.~Takeuchi, A.~Weiler and C.~Wymant,
  arXiv:1405.4295 [hep-ph].

\bibitem{sanz}
  A.~Banfi, A.~Martin and V.~Sanz,
  arXiv:1308.4771 [hep-ph].

\bibitem{azatov}
  A.~Azatov and A.~Paul,
  JHEP {\bf 1401}, 014 (2014).

\bibitem{andi}
  C.~Grojean, E.~Salvioni, M.~Schlaffer and A.~Weiler,
  arXiv:1312.3317 [hep-ph].

\bibitem{englert_spannowsky}
  C.~Englert, M.~McCullough and M.~Spannowsky,
  Phys.\ Rev.\ D {\bf 89}, 013013 (2014).

\bibitem{phi_jj}
  T.~Plehn, D.~L.~Rainwater and D.~Zeppenfeld,
  Phys.\ Rev.\ Lett.\  {\bf 88}, 051801 (2002);
  C.~Ruwiedel, N.~Wermes and M.~Schumacher,
  Eur.\ Phys.\ J.\ C {\bf 51}, 385 (2007);
 G.~Klamke and D.~Zeppenfeld,
  JHEP {\bf 0704} (2007) 052;
  K.~Hagiwara, Q.~Li and K.~Mawatari,
  JHEP {\bf 0907}, 101 (2009);
  C.~Englert, D.~Goncalves-Netto, K.~Mawatari and T.~Plehn,
  JHEP {\bf 1301}, 148 (2013);
  C.~Englert, D.~Goncalves, G.~Nail and M.~Spannowsky,
  Phys.\ Rev.\ D {\bf 88}, 013016 (2013);
  K.~Hagiwara and S.~Mukhopadhyay,
  JHEP {\bf 1305}, 019 (2013);
  M.~R.~Buckley, T.~Plehn and M.~J.~Ramsey-Musolf,
  arXiv:1403.2726 [hep-ph].

\bibitem{dieter}
  V.~Del Duca, W.~Kilgore, C.~Oleari, C.~Schmidt and D.~Zeppenfeld,
  Phys.\ Rev.\ Lett.\  {\bf 87}, 122001 (2001);
  V.~Del Duca, W.~Kilgore, C.~Oleari, C.~Schmidt and D.~Zeppenfeld,
  Nucl.\ Phys.\ B {\bf 616}, 367 (2001).

\bibitem{boos}
  E.~Boos and T.~Plehn,
  Phys.\ Rev.\ D {\bf 69}, 094005 (2004).

\bibitem{mcfm} 
  J.~M.~Campbell, R.~K.~Ellis, R.~Frederix, P.~Nason, C.~Oleari and C.~Williams,
  JHEP {\bf 1207} (2012) 092.
  J.~M.~Campbell, R.~K.~Ellis and C.~Williams, MCFM web page \url{http://mcfm.fnal.gov}

\bibitem{vbfnlo}
  F.~Campanario, M.~Kubocz and D.~Zeppenfeld,
  Phys.\ Rev.\ D {\bf 84}, 095025 (2011);
  J.~Baglio, J.~Bellm, F.~Campanario, B.~Feigl, J.~Frank, T.~Figy, M.~Kerner and L.~D.~Ninh {\it et al.},
  arXiv:1404.3940 [hep-ph].
  K.~Arnold, M.~Bahr, G.~Bozzi, F.~Campanario, C.~Englert, T.~Figy, N.~Greiner and C.~Hackstein {\it et al.},
  Comput.\ Phys.\ Commun.\  {\bf 180} (2009) 1661.

\bibitem{pythia} 
  T.~Sjostrand, S.~Mrenna and P.~Z.~Skands,
  JHEP {\bf 0605}, 026 (2006).

\bibitem{gosam} 
  J.~M.~Campbell, R.~K.~Ellis and G.~Zanderighi,
  JHEP {\bf 0610} (2006) 028;
  J.~M.~Campbell, R.~K.~Ellis and C.~Williams,
  Phys.\ Rev.\ D {\bf 81} (2010) 074023;
  H.~van Deurzen, N.~Greiner, G.~Luisoni, P.~Mastrolia, E.~Mirabella, G.~Ossola,
  T.~Peraro and J.~F.~von Soden-Fraunhofen {\it et al.},
  Phys.\ Lett.\ B {\bf 721}, 74 (2013).
     
\bibitem{powheg} 
  S.~Frixione, P.~Nason and C.~Oleari,
  JHEP {\bf 0711}, 070 (2007);
  T.~Melia, P.~Nason, R.~Rontsch and G.~Zanderighi,
  JHEP {\bf 1111}, 078 (2011);
  
\bibitem{sherpa_z}  
T.~Gleisberg, S.~.Hoeche, F.~Krauss, M.~Schonherr, S.~Schumann, F.~Siegert and J.~Winter,
  JHEP {\bf 0902}, 007 (2009)
  C.~F.~Berger, Z.~Bern, L.~J.~Dixon, F.~Febres Cordero, D.~Forde, H.~Ita, D.~A.~Kosower and D.~Maitre,
  Phys.\ Rev.\ D {\bf 78}, 036003 (2008);
  T.~Gleisberg and S.~Hoeche,
  JHEP {\bf 0812}, 039 (2008);
   S.~Hoeche, F.~Krauss, M.~Schonherr and F.~Siegert,
  JHEP {\bf 1209}, 049 (2012);
  S.~Hoche, F.~Krauss, M.~Schonherr and F.~Siegert,
  JHEP {\bf 1108}, 123 (2011);
  S.~Hoeche, F.~Krauss, M.~Schonherr and F.~Siegert,
  JHEP {\bf 1304}, 027 (2013);
  S.~Hoeche, F.~Krauss and M.~Schonherr,
  arXiv:1401.7971 [hep-ph].

\bibitem{fastjet}
  M.~Cacciari, G.~P.~Salam and G.~Soyez,
  JHEP {\bf 0804}, 063 (2008);
  M.~Cacciari, G.~P.~Salam and G.~Soyez,
  Eur.\ Phys.\ J.\ C {\bf 72}, 1896 (2012).
  
 \bibitem{fwm}
  C.~Englert, M.~Spannowsky and M.~Takeuchi,
  JHEP {\bf 1206}, 108 (2012);
  C.~Bernaciak, M.~S.~A.~Buschmann, A.~Butter and T.~Plehn,
  Phys.\ Rev.\ D {\bf 87}, 073014 (2013).

\end{thebibliography}
\end{document}